\documentclass[aps,prd,twocolumn,floatfix,nofootinbib,superscriptaddress,tightenlines]{revtex4}

\usepackage[dvipsnames]{xcolor}
\usepackage{ragged2e}
\usepackage{soul}
\usepackage{amsmath}
\usepackage{caption}
\usepackage{dcolumn}
\usepackage{lipsum}
\usepackage{amssymb}
\usepackage{url}
\usepackage{epsfig}
\usepackage{graphicx}
\usepackage{bm}
\usepackage{setspace}
\usepackage{lscape}
\usepackage{amsthm}
\usepackage{bbold}
\usepackage{dcolumn}
\usepackage{epsfig}
\usepackage{graphics}
\usepackage{graphicx}
\usepackage{subcaption}
\usepackage[justification=raggedright,singlelinecheck=false]{caption}

\usepackage{etoolbox}
\makeatletter
\patchcmd{\@makecaption}
  {\@tempdima\hfil}
  {\@tempdima\z@}
  {}{}
\makeatother
\usepackage{bm}
\usepackage{xspace}
\usepackage{cancel}
\usepackage{float}
\usepackage{multirow}
\definecolor{darkgreen}{rgb}{0,0.5,0}
\definecolor{purple}{rgb}{0.5,0,0.5}
\definecolor{nblue}{rgb}{0.0,0.0,0.50}
\definecolor{scarlet}{rgb}{1.0,0.2,0}
\definecolor{darkmagenta}{rgb}{0.55, 0.0, 0.55}
\definecolor{darkolivegreen}{rgb}{0.33, 0.42, 0.18}
\definecolor{darkcandyapplered}{rgb}{0.64, 0.0, 0.0}

\usepackage[colorlinks=true, pdfstartview=FitV, linkcolor=purple, citecolor= purple, urlcolor=blue]{hyperref}
\usepackage[normalem]{ulem}

\newcommand{\be}{\begin{equation}}
\newcommand{\tu}{\textcolor{red}{u}}

\newcommand{\fu}{\textcolor{blue}{\bar{f_2}}}
\newcommand{\fd}{\textcolor{blue}{f_1}}

\newcommand{\Meps}{\textcolor{blue}{{PS}}}
\newcommand{\Ms}{\textcolor{blue}{{S}}}

\newcommand{\Deps}{\textcolor{blue}{{DPS}}}
\newcommand{\Ds}{\textcolor{blue}{{DS}}}

\newcommand{\td}{\textcolor{darkcandyapplered}{d}}
\newcommand{\tb}{\textcolor{blue}{b}}
\newcommand{\tc}{\textcolor{darkmagenta}{c}}
\newcommand{\ts}{\textcolor{darkgreen}{s}}
\newcommand{\ee}{\end{equation}}
\newcommand{\bea}{\begin{eqnarray}}
\newcommand{\eea}{\end{eqnarray}}
\newcommand{\beas}{\begin{eqnarray*}}
\newcommand{\eeas}{\end{eqnarray*}}
\newcommand{\nn}{\nonumber}

\newcommand{\MeV}{\text{MeV}} 
\newcommand{\GeV}{\text{GeV}} 
\newcommand{\rmh}{\hat{\alpha}_{\mathrm {IR}}}

\newcommand{\eqn}[1]{Eq.~(\ref{#1})}

\newcommand{\Tr}{\text{Tr}}
\begin{document}
\title{Screening Masses for Scalar and Pseudoscalar Mesons and their Diquark Partners: Insights from the Contact Interaction Model}
\author{M. A. Ramírez-Garrido}
\email{miguelramirez.fcfm@ms.uas.edu.mx}
\affiliation{Facultad de Ciencias F\'isico-Matem\'aticas, Universidad Aut\'onoma de Sinaloa, Ciudad Universitaria, Culiac\'an, Sinaloa 80000,
M\'exico}

\author{R. J. Hern\'andez-Pinto}
\email{roger@uas.edu.mx}
\affiliation{Facultad de Ciencias F\'isico-Matem\'aticas, Universidad Aut\'onoma de Sinaloa, Ciudad Universitaria, Culiac\'an, Sinaloa 80000,
M\'exico}

\author{I. M. Higuera-Angulo}
\email{higuera@jlab.org}
\affiliation{Thomas Jefferson National Accelerator Facility, Newport News, Virginia 23606, USA}

\author{L. X. Guti\'errez-Guerrero}
\email{lxgutierrez@secihti.mx}
\affiliation{SECIHTI-Mesoamerican Centre for Theoretical Physics,
Universidad Aut\'onoma de Chiapas, Carretera Zapata Km.~4, Real
del Bosque (Ter\'an), Tuxtla Guti\'errez, Chiapas 29040, M\'exico}




\begin{abstract}
In this study, we calculate the screening masses of forty spin-zero meson and diquark states composed of light, heavy and heavy-light quarks, within a symmetry-preserving framework based on Dyson-Schwinger equations and a vector $\times$ vector contact interaction. This effective interaction reproduces hadron static properties at $T=0$ MeV that are in close agreement with experimental data and results from more sophisticated interaction kernels. 
Our analysis reveals that, at temperatures above the critical temperature, the dressed quark masses of the lightest mesons converge toward their respective current quark masses. Our work extends this study up to $T=500$ MeV.
Furthermore, we demonstrate that the screening masses of meson parity partners become degenerate above a characteristic temperature, a phenomenon that is likewise reflected in the thermal evolution of their associated Bethe-Salpeter amplitudes (BSAs), providing robust evidence for chiral symmetry restoration. A similar trend is observed for scalar and pseudoscalar diquarks, whose screening masses also converge at high temperatures. As a consequence, these diquarks shall contribute equally to the baryon screening masses in the high-temperature regime.
Wherever results from other approaches are available, we perform direct comparisons and observe consistency with our findings.
\end{abstract}


\maketitle

\section{Introduction}
The recreation of the quark-gluon plasma (QGP) existed in the early stages of the universe is a key objective of current experiments with heavy-ion collisions \cite{Ayala:2016vnt,Elfner:2022iae}.
These collisions can reproduce similar conditions  to those existed in the early universe. The QGP state arises when bound states are deconfined at high temperatures and chiral symmetry is restored. In this hot hadronic matter a chiral phase transition is believed to occur at a certain critical temperature $T_c$, which has been estimated to be around 
$T_c=155$ MeV \cite{Andronic:2017pug,Bazavov:2011nk}.
The QGP can shed light on strong interactions using non-zero temperatures, and on open questions regarding confinement, hadronization, and dynamical chiral symmetry breaking, as well as the evolution of matter in the early universe \cite{Yagi:2005yb}.

The formation of QGP cannot be observed directly; it must be inferred through the properties of the hadrons that are formed in the collisions. 
As the temperature increases, the theory transitions from a confined phase with hadronic degrees of freedom, where chiral symmetry is spontaneously broken, to a deconfined phase of quarks and gluons, where chiral symmetry is restored.
This process takes place during the plasma phase, and the key element to understand this restoration is precisely the pion, which is the boson associated with the dynamical breaking of this symmetry.  Consequently, investigating the thermal evolution of pion properties is relevant to the Hadron Physics program. 
On the other hand, mesons composed of light quarks are not the only relevant states in this context.
Heavy-quark bound states, commonly known as quarkonia, have garnered significant interest in recent years \cite{Yao:2018zrg,Rapp:2009my}. Experimental results from collaborations at RHIC, SPS, and the LHC \cite{STAR:2005gfr,ALICE:2013osk,NA50:2004sgj,CMS:2012gvv} have shown that these states are powerful probes for studying  the QGP. This experimental focus has been complemented by numerous theoretical investigations \cite{Veliev:2010gb, Aarts:2022krz,Bazavov:2020teh,Mukherjee:2008tr,Veliev:2011zz}.\\
At finite temperature, the notion of mass requires careful consideration, as two distinct definitions emerge: the pole mass and the screening mass. The pole mass is identified through the pole of the propagator, while the screening mass is derived from the behavior of hadronic correlation functions at large spatial distances. 
The mass dependence of hadrons as a function of temperature has been studied using various methods, including lattice QCD \cite{Aarts:2015mma,Giusti:2024ohu,Takeuchi:2017wii}.
In particular, in the meson sector \cite{Rapp:1999ej,Maris:2000ig,Wang:2013wk,Dominguez:2012um,Sumit:2023hjj,Mallik:2002ef,Dominguez:1996kf,Song:1993ipa},
it is well known that as the temperature approaches infinity, bound states dissolve, and consequently, meson masses approach their free-theory limit of $2\sqrt{(\pi T)^2 +m_q^2 }$, with $m_q$ being the quark mass, independent of its spin-parity structure \cite{Florkowski:1993bq,Bazavov:2014cta,Karsch:2012na}. At extremely high temperatures and in the limit of vanishing quark masses, the resulting behavior will be $2\pi T$
\cite{Eletsky:1988an,Alberico:2007yj}.
 When symmetry is restored, the screening masses of meson parity partners become degenerate.
In relativistic quantum field theory, the parity partner of a given state can be generated through a simple chiral rotation of that state. If chiral symmetry were exact, parity partners would have identical masses. Nonetheless, the observed spectra of bound states clearly deviate from this symmetry. One of the most striking signatures of chiral symmetry restoration is the emergence of degenerate screening masses between parity partners.
Pseudoscalar diquarks are anticipated to play a role as significant as that of scalar diquarks.

\begin{figure}[h]
   \vspace{-4cm}
   \centering
    \includegraphics[scale=0.4,angle=0]{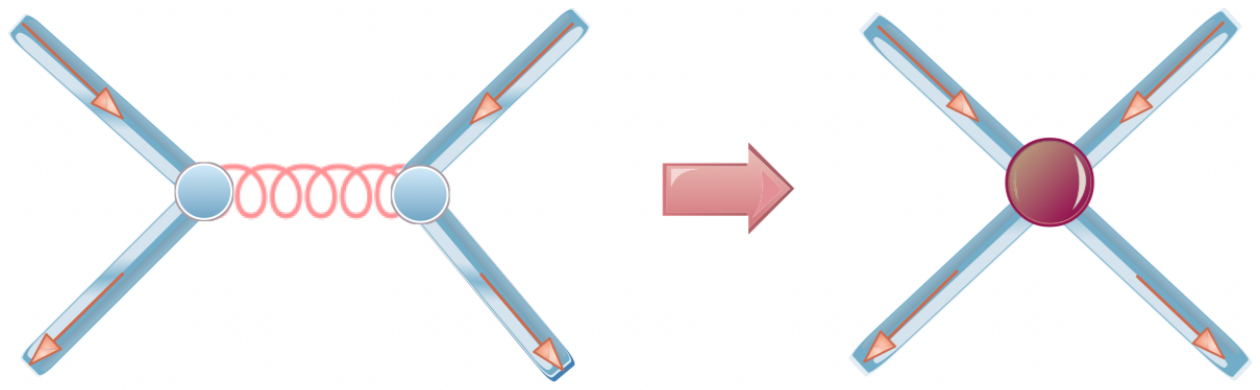}
    \vspace{-4cm}
    \caption{\justifying Diagrammatic illustration of the Contact Interaction, constructed from the simplified gluon propagator model introduced in Eq.~\eqref{eqn:contact_interaction}.}
    \label{fig:ci}
\end{figure}

Results for screening masses of mesons built from light and strange quarks in the
temperature range of approximately between 140 MeV to 800 MeV can be found in Ref.~\cite{Cheng:2010fe}. Besides, the dependence of bottomonium masses on temperature is studied in Refs.~\cite{Petreczky:2021zmz,Rothkopf:2019ipj}. Meson screening masses for different channels in $\ts \bar{\ts}$, $\ts \bar{\tc}$ and
$\tc \bar{\tc}$ sectors are found in Ref.~\cite{Kaczmarek:2022oiu}. For charmonia see Ref.~\cite{Navarra:1995gv}. The higher mass of heavy quarks allows them to remain stable at higher temperatures compared to mesons formed by light quarks.
In this paper, we explore the screening masses of mesons composed of light and heavy quarks using the Contact Interaction (CI) model formalism \cite{Gutierrez-Guerrero:2010waf}. Within this framework, the gluon propagator becomes momentum-independent Fig.~\ref{fig:ci}; see Ref. \cite{GutierrezGuerrero:2010md}  for details. 

Since its earliest predictions, this model has proven to be an excellent tool for computing hadronic observables and a reliable guide for calculations of meson and baryon form factors \cite{Gutierrez-Guerrero:2019uwa,Gutierrez-Guerrero:2021rsx, Gutierrez-Guerrero:2024him,Hernandez-Pinto:2024kwg,Paredes-Torres:2024mnz}.
The first calculations of screening masses for light mesons and baryons in the ground state using CI formalism at nonzero temperature were presented in \cite{Wang:2013wk}, and more recently in \cite{Chen:2024emt}, where mesons containing the strange quark were included. In addition, the CI model has also been used to investigate the QCD phase diagram under the influence of an external magnetic field $eB$, at finite temperature $T$ and quark chemical potential $\mu$ \cite{Ahmad:2025prr}.

Our work is organized as follows: in Section \ref{CI}, we describe the main features of our model, including the gap equation, which we study for different quark flavors. In Section \ref{Bse-av}, we incorporate temperature into the Bethe–Salpeter equation and investigate the screening masses of pseudoscalar (PS) and scalar (S) mesons, as well as their corresponding diquark partners. Finally, in Section \ref{Summary}, we present our summary and outlook.
\section{Contact interaction: key features} \label{CI}
In this section, we provide a concise overview of the CI model, highlighting the fundamental elements required for the computation of screening masses of mesons and diquarks. This section includes the gap equation, the model parameters adopted, the temperature dependence of the quark masses, and the notation that is consistently used throughout our analysis.
\subsection{The Gap Equation at Finite Temperature}
In analogy with the temperature-independent case, the core of our study for $T\neq 0$ lies in the dressed-quark propagator for a quark of flavor $f$, which is determined by solving the gap equation,
\begin{align}
    \nn S_f^{-1}(p;T) &=i \vec{\gamma} \cdot \vec{p} +i \gamma_4 \omega_n +m_f \\
    &+ \frac{16 \pi \hat{\alpha}_{\mathrm{IR}}}{3} \int_{\ell,dq} \gamma_{\mu} S_f(q;T)\gamma_{\mu}\,,
\end{align}
where $p=(\omega_n,\vec{p})$ and $q=(\omega_\ell,\vec{q})$ are the four momenta of the quarks, $f$ represents the quark flavour, $m_f$ the current mass of the quark, and $\omega_n =(2n+1)\,\pi \,T$ is the fermion Matsubara frequency.
Following Ref.~\cite{Chen:2024emt}, we have used the notation,
\bea\int_{\ell,dq}:=T\sum^{\infty}_{\ell=-\infty}\int \frac{d^3\vec{q}}{(2\pi)^3}\,.\eea
For the gluon propagator, we have used the  CI approach, i.e., 
\begin{eqnarray}
\label{eqn:contact_interaction}
g^{2}D_{\mu \nu}(k)&\to&4\pi\,\hat{\alpha}_{\mathrm{IR}}\,\delta_{\mu \nu}  \,,
\end{eqnarray}
\noindent  where $\hat{\alpha}_{\mathrm{IR}}=\alpha_{\mathrm{IR}}/m_g^2$, with $\alpha_{\mathrm{IR}}$ the running-coupling and \,$m_g=500\,\MeV$  the gluon mass scale~\cite{Aguilar:2017dco,Binosi:2017rwj,Gao:2017uox}. 
The dressed mass $M_f(T)$ is then studied by
\begin{align}
\label{dressed mass}
    M_f(T) =m_f +\frac{16 \pi \hat{\alpha}_{\mathrm{IR}}}{3} \int_{\ell,dq} \frac{4M_f(T)}{s_{\ell}+M_f(T)^2}  \, ,
\end{align}
    with $s_\ell=\vec{q}+\omega_\ell$, and using the Poincaré invariant method, we find,
    \begin{align}\label{gap-T}
        M_f(T) =m_f +M_f(T)\frac{4  \hat{\alpha}_{\mathrm{IR}}}{3\pi} \mathcal{C}^{\rm iu} (M_f(T)^2;T) 
    \end{align}
    where
    \begin{align}
        \nn \mathcal{C}^{\rm iu} (M_f(T)^2;T) = 2 \sqrt{\pi} \, T \, \int_{\tau_{\rm UV}^2}^{\tau_{\rm IR}^2} d\tau \, \frac{e^{- M_f^2 \, \tau} \vartheta_2\left(e^{- 4\pi^2 T^2\, \tau}\right)}{\tau^{3/2}}\,,
    \end{align}
with $\vartheta_2(x)$ the Jacobi theta function
and 
${\cal C}^{\rm iu}(\sigma;T)/\sigma = \overline{\cal C}^{\rm iu}(\sigma;T)\,$.
$\tau_{\rm IR, UV}$ are respectively, infrared and ultraviolet regulators or, similarly, $\Lambda_{\rm IR,UV} = \tau_{\rm IR, UV}^{-1}$.  
\subsection{Parameters}
For our calculations, we adopt the parameters introduced in Ref.~\cite{Gutierrez-Guerrero:2021rsx}, where a unified framework was employed to compute the masses of mesons and baryons composed of both light and heavy quarks in CI model. These parameters were also used to describe successfully the parity partners.
Therefore, we use the parameters listed in Table~ \ref{parameters}.
 \begin{table}[H]
 \caption{ \justifying \label{parameters} 
 Parameters used for the ultraviolet regulator and the coupling constant for different quark combinations  $\hat{\alpha}_{\mathrm {IR}}=\hat{\alpha}_{\mathrm{IRL}}/Z_H$, where $\hat{\alpha}_{\mathrm {IRL}}=4.57$. $\Lambda_{\rm IR}^{(0)} = 0.24$ GeV is a fixed parameter at $T=0$ MeV; for nonzero temperatures this parameter is modified according to Eq. (\ref{irt}).}
\begin{center}
\label{parameters1}
\begin{tabular}{@{\extracolsep{0.0 cm}} || l | c | c | c ||}
\hline \hline
 \, quarks \, &\,  $Z_{H}$ \, &\,  $\Lambda_{\mathrm {UV}}\,[\GeV] $ \,  &\,  $\hat{\alpha}_{\mathrm {IR}}$
 \\
 \hline
 \rule{0ex}{2.5ex}
$\, \tu,\td, \ts$ & 1 & 0.905 & 4.57   \\ 
\rule{0ex}{2.5ex}
$\, \tc, \td, \ts $ & \, 3.034 \, & 1.322 & 1.51 \\ 
\rule{0ex}{2.5ex}
$\, \tc $ & \, 13.122 \, & 2.305 & 0.35 \\ 
\rule{0ex}{2.5ex}
$\,  \tb,\tu$, \ts & \, 16.473 \, & 2.522 & 0.28 \\ 
\rule{0ex}{2.5ex}
$\, \tb, \tc$     &  59.056 & 4.131 & 0.08 \\
\rule{0ex}{2.5ex}
$\, \tb $ & 165.848 & 6.559 & 0.03\\
\hline \hline
\end{tabular}
\end{center}
\end{table}
The values in the Table \ref{parameters} indicate that $\hat{\alpha}_{\mathrm {IR}}$ naturally varies with the mass scale $\Lambda_{\mathrm{UV}}$ as described in Ref.~\cite{Bedolla:2015mpa} and subsequently employed in various later studies~\cite{Bedolla:2016yxq, Raya:2017ggu, Gutierrez-Guerrero:2019uwa, Yin:2019bxe, Yin:2021uom}. 
It is important to highlight that, in the calculation of hadron masses in their ground and excited states  at $T=0$ MeV, we used a constant $\Lambda_{\rm IR} = 0.24$ GeV, since this infrared regulator implements confinement by
ensuring the absence of quark production thresholds in
all processes. However, for screening masses, this value must vary with temperature \cite{Chen:2024emt,Wang:2013wk}, that is,
$\Lambda_{\rm IR} \to\Lambda_{\rm IR}(T)\,.$
We followed previous works \cite{Chen:2024emt,Wang:2013wk} and defined  the infrarred regulator evolution as,
\bea
\label{irt}
\Lambda_{\rm IR}(T) =\Lambda_{\rm IR}^{(0)} \left(\frac{ M_f(T)}{M_f(0)}\right)^{1/4} \, ,
\eea
 where the value of $\Lambda_{\rm IR}^{(0)}$
  corresponds to the constant infrared scale used at $T=0$ MeV, namely $0.24$ GeV.
With all these parameters in hand, the following subsection explores the behavior of the dressed quark masses for the $\tu$, $\ts$, $\tc$, and $\tb$ flavors\footnote{We include $\td$-quarks by imposing isospin symmetry.} at nonzero temperatures. Where applicable, our results are compared with those obtained from more sophisticated approaches.
\subsection{Quark Masses}
\begin{figure}[t]
    \centering    \includegraphics[width=1\linewidth]{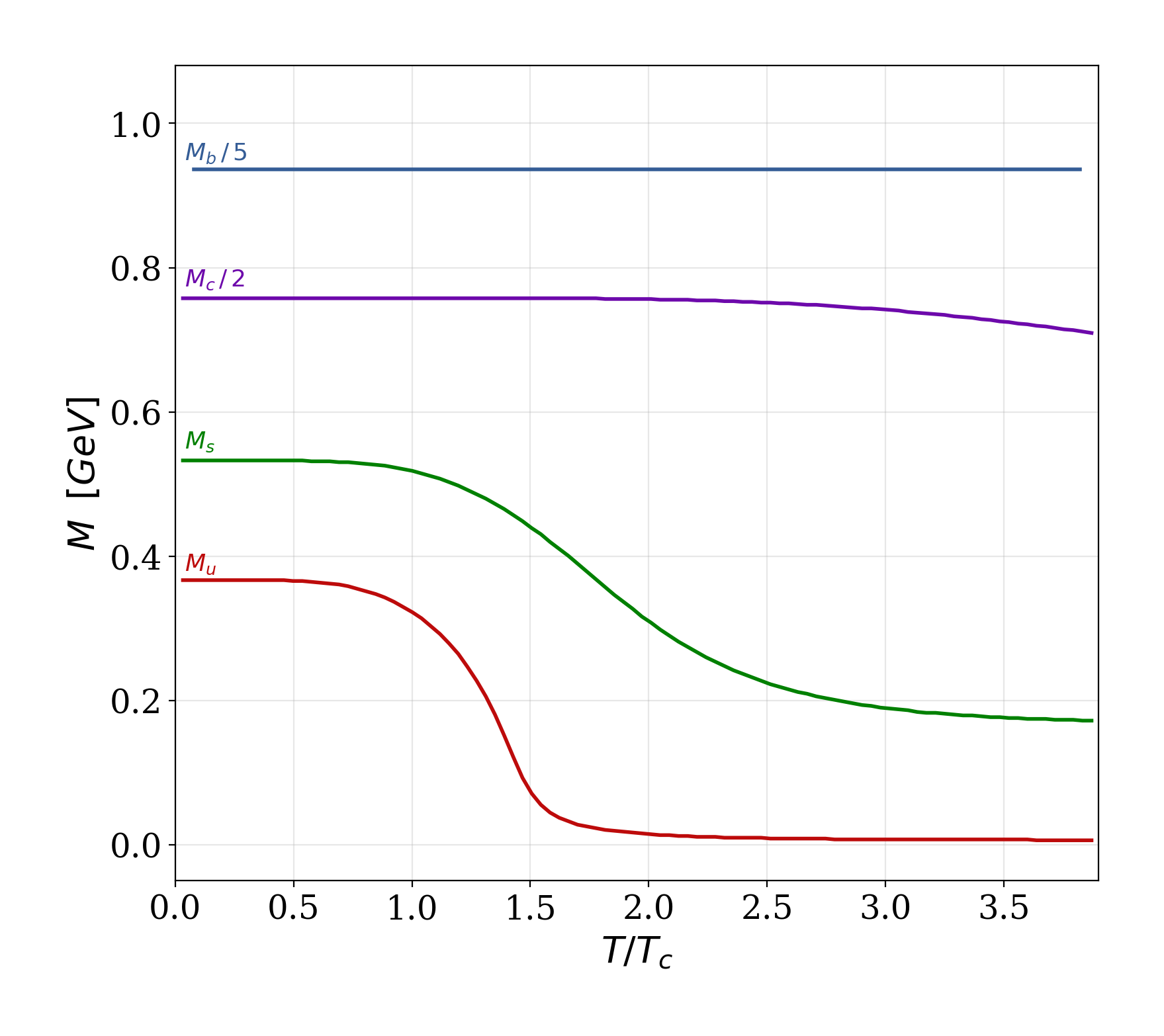}
    \vspace{-1cm}
    \caption{\justifying Results for the screening quark masses computed across a temperature range from $0$ to $4\, T_c$, with $T_c=155$ MeV \cite{Andronic:2017pug,Bazavov:2011nk}. At high temperatures, the quark mass asymptotically tends toward its bare mass, $M(T)\to m$.}
    \label{fig:dressed}
\end{figure}

Table \ref{table-M} establishes the current quark masses used in our calculations at $T=0$ MeV; however, we now compute the dressed quark masses as a function of the temperature using the parameters presented above.
Table~\ref{table-M} presents the results of our calculation using \eqn{dressed mass} for the dressed  mass at zero temperature. \\
\begin{table}[h!]
\caption{\justifying \label{table-M}
Current ($m_{f}$) and dressed masses
($M_{f}$) for quarks in GeV obtained using $T=0$ MeV.}
\begin{center}
\begin{tabular}{@{\extracolsep{0.0 cm}} || c | c | c | c || }
\hline 
\hline
 $m_{\tu}=0.007$ &$m_{\ts}=0.17$ & $m_{\tc}=1.08$ & $m_{\tb}=3.92$   \\
 \rule{0ex}{2.5ex}
 $ M_{\tu}=0.367$ \, & \, $  M_{\ts}=0.53$\; \, &\,   $  M_{\tc}=1.52$ \, &\,  $  M_{\tb}=4.75$   \\
 \hline
 \hline
\end{tabular}
\end{center}
\end{table}
We illustrate the evolution of the screening masses of these quarks as a function of temperature in Fig.~\ref{fig:dressed}.
In the present work, which includes mesons containing heavy quarks, we adopt $T_c = 155$ MeV, consistent with values used in lattice QCD simulations \cite{Andronic:2017pug, Bazavov:2011nk}, to facilitate comparison with other approaches. However, it is important to emphasize that this temperature acts merely as a rescaling factor in our approach.

\begin{figure}[h]
    \vspace{-0.5cm}
    \centering
    \includegraphics[width=1\linewidth]{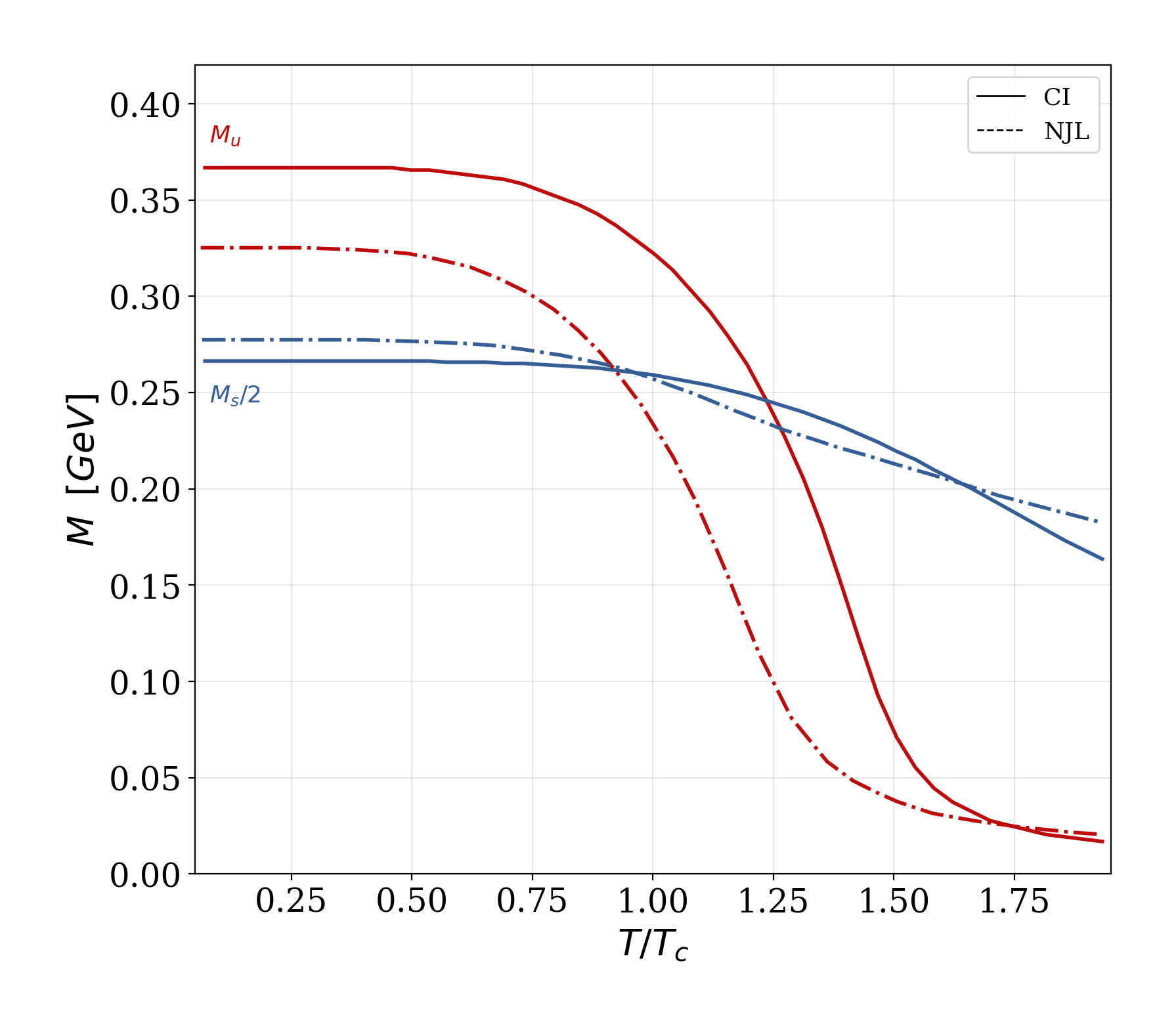}
    \vspace{-1cm}
    \caption{\justifying Quark dressed masses obtained here (solid lines) and those derived from the NLJ model (dashed lines) in Ref.~\cite{Sheng:2022ssp}.}
    \label{fig:compa-njl}
\end{figure}

It is worth highlighting that, as the temperature rises, the quark mass decreases, a trend that is particularly pronounced for light quarks.
We make this observation more explicit in Table~\ref{T600}, where the current and constituent (dressed) quark masses are compared at $T=500$ MeV. The largest relative deviation is observed for the charm quark, reaching 29\% with respect to the screening mass, followed by the bottom quark with a deviation of 17.5\%. For lighter quarks, the discrepancies remain below 3\%.
\begin{table}[t]
    \centering
    \caption{\justifying Screening masses of the quarks at $T=500$ MeV in comparison with current masses. All masses are in GeV.}
    \label{T600}
    \begin{tabular}{@{\extracolsep{0.0 cm}}c|c|c}
    \hline\hline
    \, Quark \, & \, $m_f^{\rm CI}$ \, & \, $M(T=500\text{ MeV})$ \, \\
    \hline
       $\tu$, $\td$   & 0.007 & 0.007 \\
       $\ts$   &  0.17& 0.176\\
       $\tc$    &  1.08& 1.447\\
       $\tb$  &  3.92 & 4.676\\ 
       \hline\hline 
    \end{tabular}
\end{table}
In addition, we point out that the behavior of the  dressed quark masses as a function of temperature is as expected. For instance, for the $\tu$ and $\ts$ quarks, we compare our results with those obtained using the NJL model \cite{Sheng:2022ssp} in Fig. \ref{fig:compa-njl} showing the correct trend.\\
%
%
%
%
\section{Bethe Salpeter Equation}
\label{Bse-av}
\begin{figure}[H]
   \centering
    \includegraphics[scale=0.45,angle=0]{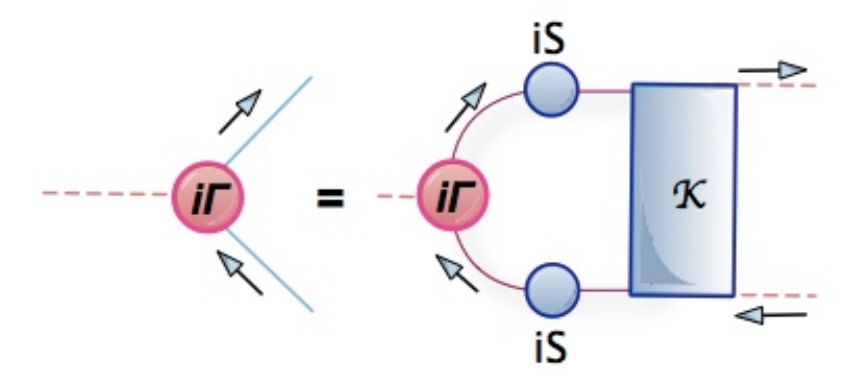}
    \caption{\justifying Diagrammatic representation of the BSE. Blue (solid) circles represent dressed quark propagators $S$, red (solid) circle is the meson BSA $\Gamma$ while the blue (solid) rectangle is the dressed-quark-antiquark scattering kernel ${\mathcal {K}}$.}
    \label{fig:BSEfig}
\end{figure}

Hadrons formed by two quarks, including mesons and diquarks, can be described as relativistic bound states using the homogeneous Bethe–Salpeter equation (BSE), as illustrated diagrammatically in Fig.~\ref{fig:BSEfig}. \\
 This equation, in the case of nonzero temperatures, can be mathematically expressed as~\cite{Salpeter:1951sz},
 \begin{equation}
[\Gamma(Q;T)]_{tu} = \int \! \frac{d^4q}{(2\pi)^4} [\chi(Q;T)]_{sr} {\mathcal K}_{tu}^{rs}(Q;T)\,,
\label{genbse}
\end{equation}
where $[\Gamma(Q;T)]_{tu}$ represents the bound-state's Bethe Salpeter Amplitude (BSA) and $\chi(Q;T) = S(q+Q;T)\Gamma(Q;T) S(q;T)$ is the Bethe Salpeter wave-function; $r,s,t,u$ represent colour, flavor and spinor indices; and ${\mathcal K}$ is the relevant quark-antiquark scattering kernel. Additionally, $Q=\{\vec{Q},\Omega_m\}$ with $\Omega_m=2m\pi T$ is the Boson Matusbara frequency. Throughout this article, our analysis is restricted to the zeroth Matsubara frequency.
A general decomposition of the BSA for mesons ($\fd\fu$) in the CI is given by, 
\begin{align}
    \Gamma_i(Q;T)= E_i(T) \, A_i + F_i(T) \, B_i\,,
\end{align}
with $i$ denoting the type of hadron under study, specifically S and PS mesons and diquarks; in the following, we shall discuss the explicit analytical expressions for each case analyzed in this work. These channels play a central role in the investigation of chiral symmetry restoration, as will be further elaborated below.
\subsection{Scalar Mesons}
For our analysis we consider mesons composed of heavy ($Q\bar{Q}$), heavy-light ($Q\bar{q}$), and light ($q\bar{q}$) mesons, focusing on the flavors $\tu$, $\td$, $\ts$, $\tc$, and $\tb$, with the $SU(5)$ multiplets predicted by the quark model~\cite{GellMann:1964nj,Zweig:1964jf,Zweig:1981pd}.

\begin{figure}[H]`
       \centerline{
       \includegraphics[scale=0.25,angle=0]{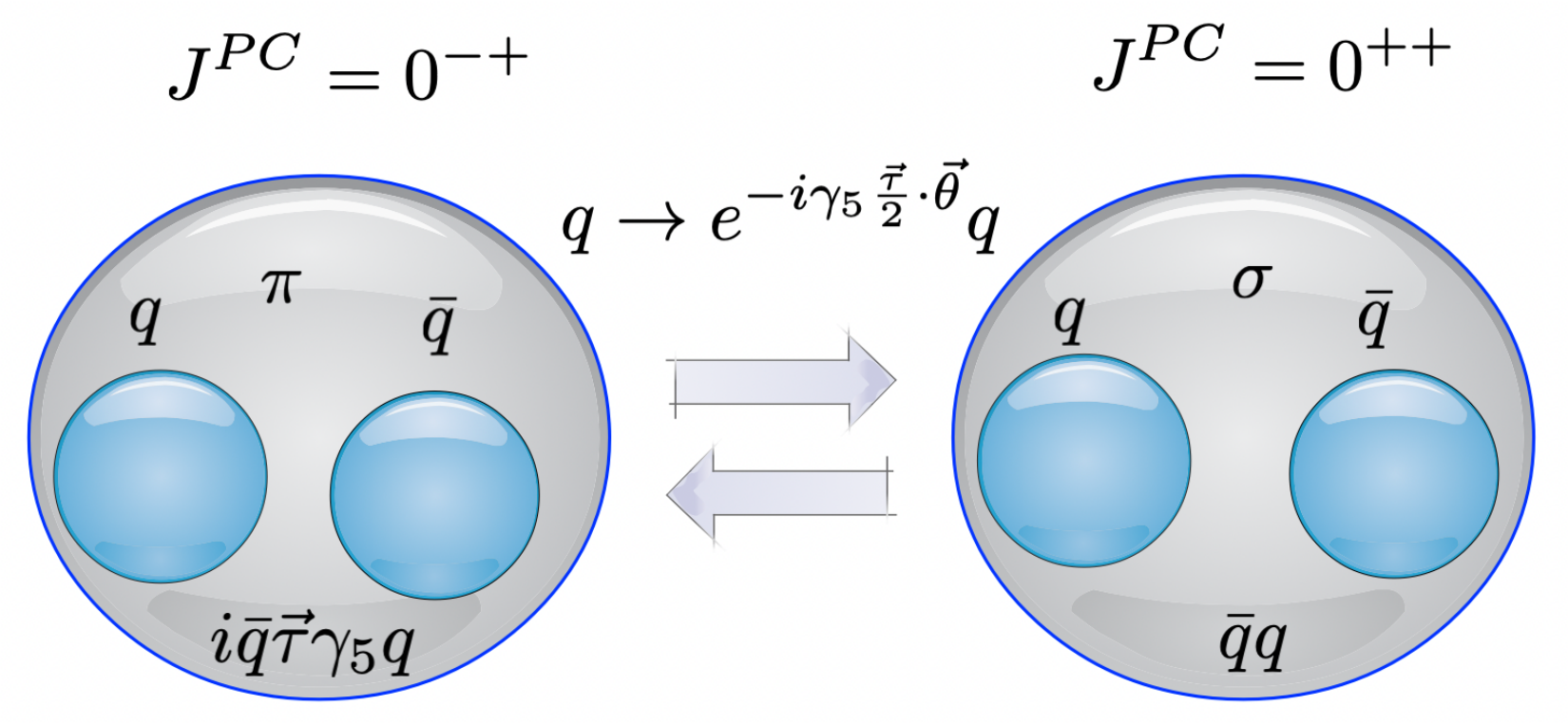}
       }
       \caption{\justifying \label{sca-spi}The S meson is viewed as the parity partner of PS. Note that the S in this article only refer to their quark-antiquark content.} 
\end{figure}

We begin with S mesons, which are of great importance, as they are the chiral partners of the PS mesons; that is, they can be obtained through a chiral transformation \cite{Hernandez-Pinto:2023yin}, as illustrated in Fig. \ref{sca-spi}. 
\begin{figure*}[t!]
\begin{tabular}{@{\extracolsep{-2.3 cm}}c}
 \renewcommand{\arraystretch}{-1.6} %
 \hspace{-1cm}
 \includegraphics[scale=0.6]{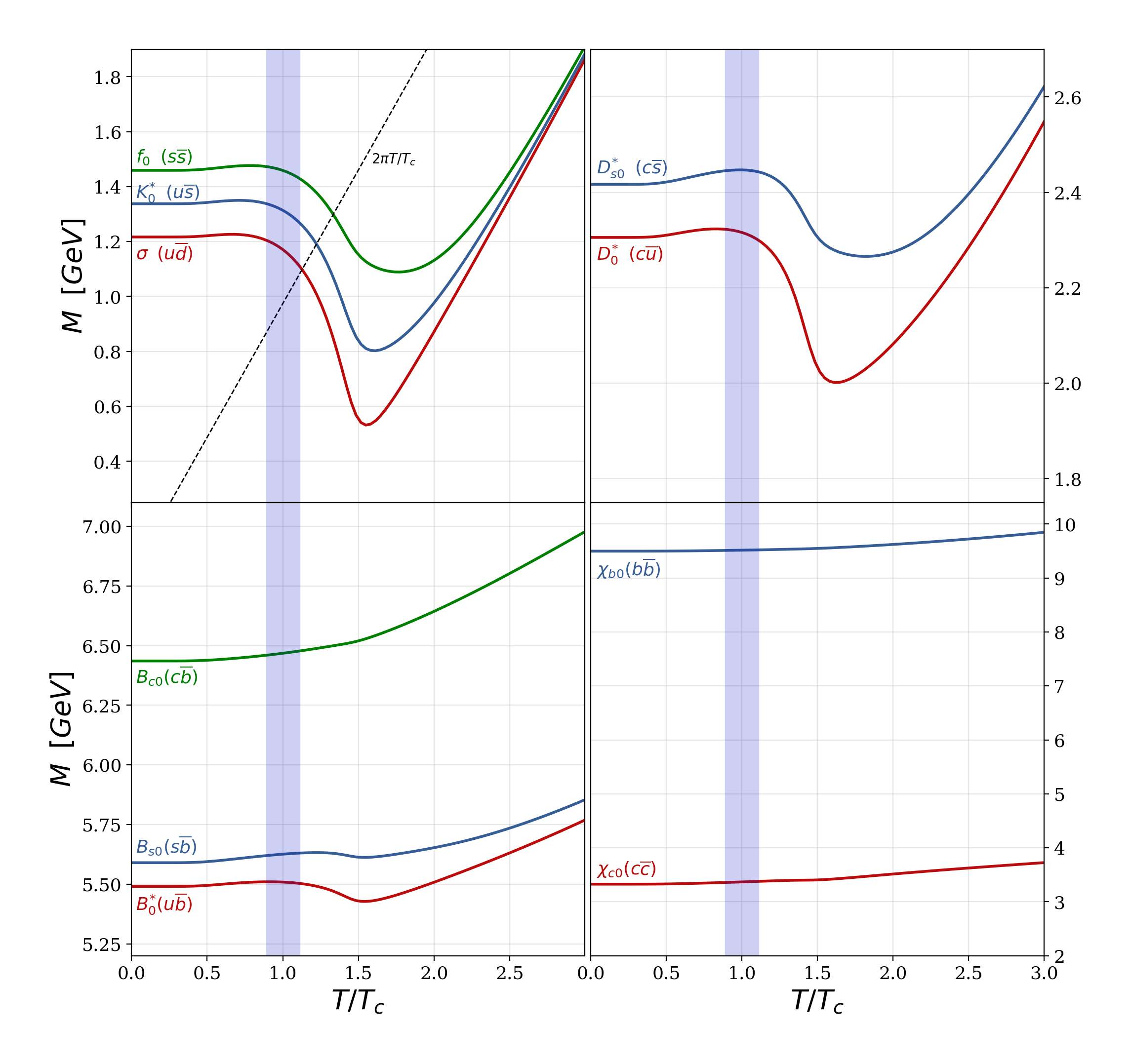}
\end{tabular}
\vspace{-1cm}
\caption{\justifying \label{fig:sscalar}The screening mass for S mesons composed of the five quark flavors ($\tu$, $\td$, $\ts$, $\tc$ and $\tb$). 
In the upper left panel, the lightest mesons are shown; the upper right  panel displays mesons composed of one light quark and one charm quark. The lower  left panel shows mesons containing different quarks, including a bottom quark, while the lower right panel presents mesons composed of two identical heavy quarks. The shaded purple band indicates the critical temperature $T_c\pm  0.1 \, T_c$. We can observe that near this value most of the screening masses exhibit a change in their behavior.
}
\end{figure*}
With a symmetry-preserving regularization of the interaction, the Bethe-Salpeter amplitude is independent of the relative momentum, and the BSA for a S meson comprised of quarks with flavor $\fd$ and antiquarks with flavor $\fu$ at nonzero temperature can be expressed as:
\begin{equation}
\label{sBSA}
\Gamma_{\Ms}(Q;T) = \mbox{\boldmath $I$}_{\rm D} \, E_{\Ms}(T)\,,
\end{equation}
which corresponds to the eigenvalue equation:
\bea \label{eigenS}
1\hspace{-0.06cm}+\hspace{-0.06cm}{\cal K}_{S}(Q^2; T)\hspace{-0.06cm}=\hspace{-0.06cm}0,
\eea
where
\begin{eqnarray}\nn
\nonumber
{\cal K}_{S}(Q^2;T) &=&
- \frac{4\rmh}{3\pi}
\int_0^1d\alpha\,\bigg[-{\cal L}_G (Q;T)
\overline{\mathcal{C}}_1^\mathcal{T}\\
 &+&\bigg(\mathcal{C}_1^\mathcal{T}
-\overline{\mathcal{C}}_1^\mathcal{T}\bigg) \bigg], \label{Kse}
\end{eqnarray}
and
\bea  \hspace{-4mm}\nn {\cal L}_{G}(Q;T)&=&M_{\fd} M_{\fu}+\alpha (1-\alpha)Q^2, \\
   \nn \mathcal{C}^\mathcal{T}&:=& \mathcal{C}^{\rm iu}(\omega^{(1)}(T);T)\,,\\
  \nn \overline{\mathcal{C}}_1^\mathcal{T}&:=& \overline{\mathcal{C}}_{1}^{\rm iu}(\omega^{(1)}(T);T)\,,
\end{eqnarray}
with
\begin{eqnarray} 
 \nn   \omega^{(1)}(T)&=&M_{\fu}^2(T)(1-\alpha)+\alpha M_{\fd}^2(T) +\alpha(1-\alpha)Q^2,\\
    \nn
    \mathcal{C}_{1}^{\rm iu}(z;T)&=&-z(d/dz)\mathcal{C}^{\rm iu}(z;T)\,.
\end{eqnarray}
The  \eqn{eigenS} has a solution when $Q^2=-m_{\Ms}^2$ and the canonical normalisation condition is given by
\begin{equation}
\frac{1}{E_{\Ms}^2(T)} = - \left. \frac{9}{2} m_G^2 \frac{d}{dQ^2} \mathcal{K}_{\Ms}(Q^2;T)\right|_{Q^2=-m_{\Ms}^2}.\;
\end{equation}
As is well known, dynamical chiral symmetry breaking generates a large dressed-quark anomalous chromomagnetic moment and, as a result, the spin-orbit splitting between ground-state mesons and their parity partners is significantly enhanced~\cite{Bermudez:2017bpx,Bashir:2011dp,Chang:2010hb,Chang:2011ei}. In this context, for nonzero temperatures, we capture this behavior and follow the approach outlined in Ref.~\cite{Roberts:2011cf}, introducing spin-orbit repulsion into the S meson channel through the phenomenological coupling $g_{\rm SO} \leq 1$. It is evident that, for the screening masses this factor, which was previously taken as a constant parameter, must now depend on temperature, as follows: 
\begin{equation}
\label{gsom}
 g_{\rm SO}^{q\bar{q},0^+}(T) = 1-\frac{M_u(T)}{M_u(0)}(1-{0.32}^2)\,,
\end{equation}
such that for $T=0$ MeV, one recovers the static value $g_{\rm SO}^{q\bar{q},0^+}(0)= {0.32}^2$.

\begin{table}[H]
\caption{\justifying \label{table-mesones-esc}
Computed and experimental masses for S mesons (GeV) and BSA in $T=0$ MeV. Column five shows the reduced mass for each meson, and column six shows the limit
$2\sqrt{(\pi T)^2 +(2 M_R)^2 }$ at $T=500$ MeV. The relative increase from $T=0$ MeV to $T=500$ MeV
in the free limit is approximately 200\% for the mass of the $\sigma$ meson.The mass of the $\sigma$-meson, marked with an asterisk, denotes the estimate for the state’s dressed-quark core \cite{Pelaez:2006nj,RuizdeElvira:2010cs}.} 
\begin{center}
\begin{tabular}{@{\extracolsep{0.3 cm}}ccccc|c}
\hline
\hline
Meson   &Exp.& CI & $E_{\Ms}$ & $2M_R$ & Limit\,\\ \hline 
\rule{0ex}{2.5ex}
$\sigma(\tu\bar{\td})$ & 1$-1.2^{\ast}$ &1.22&0.66 & 0.367 & 3.61\\
\rule{0ex}{2.5ex}
$K_0^*(\tu\bar{\ts})$ & $\cdots$ & 1.34 &0.65&0.433 & 3.63\\
\rule{0ex}{2.5ex}
$f_0(\ts\bar{\ts})$ & $\cdots$ & 1.46 &0.64& 0.53 &3.66\\
\rule{0ex}{2.5ex}
$D_0^*(\tc\bar{\tu})$&2.300&2.31&0.39& 0.591 &3.68\\
\rule{0ex}{2.5ex}
$D_{s0}^*(\tc\bar{\ts})$& 2.317&2.42&0.37&0.785& 3.76\\
\rule{0ex}{2.5ex}
$B_{0}^*(\tu\bar{\tb})$& $\cdots$ &5.49&0.21&0.681& 3.71\\
\rule{0ex}{2.5ex}
$B_{s0}(\ts\bar{\tb})$& $\cdots$ &5.59&0.20&0.953 &3.83\\
\rule{0ex}{2.5ex}
$B_{c0}(\tc\bar{\tb})$& $\cdots$ &6.44&0.08& 2.30 &4.78\\
\rule{0ex}{2.5ex}
$\chi_{c0}(\tc\bar{\tc})$&3.414&3.33&0.16&1.52& 4.18\\
\rule{0ex}{2.5ex}
$\chi_{b0}(\tb\bar{\tb})$ &9.859&9.50&0.04 &4.75 &7.00\\
\hline
\hline
\end{tabular}
\end{center}
\end{table}

With all the parameters shown above, we present our results for the masses at 
$T=0$ MeV in Table \ref{table-mesones-esc}. 
For our calculations involving mesons composed of different quark flavors, we modify \eqn{irt} by replacing $M_f$ with the reduced mass $M_R = M_{\fd} M_{\fu}/[M_{\fd} + M_{\fu}]$.
It is clear that at zero temperature, and for mesons with masses above 1.5 GeV, our results can be directly compared with experimental data \cite{ParticleDataGroup:2024cfk}, showing good agreement. However, mesons with lower masses such as the $\sigma(\tu\bar{\td})$,  $K_0^*(\tu\bar{\ts})$ and 
$f_0(\ts\bar{\ts})$, represent special cases due to ongoing controversy regarding their internal structure \cite{Santowsky:2020pwd,Santowsky:2021ugd,Santowsky:2021lyc}. For instance, the $\sigma$-meson does not predominantly exhibit a conventional $q\bar{q}$
  structure. Nonetheless,  unitarised chiral perturbation theory indicates the presence of a subdominant $q\bar{q}$ component emerging at a mass of approximately 1$-$1.2 GeV  \cite{Pelaez:2006nj}. It is this $q\bar{q}$ component of the $\sigma$-mesons to which we refer in Table \ref{table-mesones-esc}, while the $K_0^*(\tu\bar{\ts})$ and $f_0(\ts\bar{\ts})$ scalar mesons are considered hypothetical states in the context of our analysis, included primarily for completeness and to enable the construction of their corresponding diquark partners in future baryonic studies.

Fig.  \ref{fig:sscalar} displays the screening masses of the S mesons. 
Their screening masses remain nearly constant for $T\approx T_c/2$. As the temperature increases, the screening masses decrease while, as we shall see, they still remain larger than those of their corresponding parity partners—until they reach a minimum which is largely attributed to the temperature dependence of 
$g_{\rm SO}^{q\bar{q},0^+}(T)$ in \eqn{gsom}.  Beyond this point, they start to rise again. This trend observed for the S mesons agrees with previous findings reported in Refs.
\cite{Fu:2009zs,Bhattacharyya:2019qhm,Li:2019nzj}.
In Table~\ref{temps scalar}, we report the temperature $T_H$ at which this turnaround occurs, pointing out the onset of accelerated growth.
 \begin{table}[htbp]
 \caption{ \justifying \label{temps scalar} 
 Minimum temperatures $T_H$ above which the S meson masses begins to increase. All quantities are in GeV.} 
\begin{center}
\label{temp1}
\begin{tabular}{@{\extracolsep{0.0 cm}}  c | c | c | c }
\hline \hline
 \, Meson \, &\,  $T_{H}$ \, & \, Meson \, &\,  $T_{H}$ \, 
 \\
 \hline
 \rule{0ex}{2.5ex}
$\, \sigma(\tu\overline{\td})$ & 0.241 & $\, B_0^{*}(\tu\overline{\tb}) $ & \, 0.239 \,\\ 
\rule{0ex}{2.5ex}
$\,  K_0^{*}(\tu\overline{\ts}) $ & \, 0.248 \, & $\, B_{s0}(\ts\overline{\tb}) $ & 0.239 \\ 
\rule{0ex}{2.5ex}
$\,  f_0(\ts\overline{\ts}) $ & \, 0.272 \, & $\, B_{c0}(\tc\overline{\tb}) $ &0.075\\ 
\rule{0ex}{2.5ex}
$\,   D_0^{*}(\tc\overline{\tu})$ & \, 0.252 \, & $\, \chi_{c0}(\tc\overline{\tc}) $ & 0.085 \\ 
\rule{0ex}{2.5ex}
$\,  D_{s0}^{*}(\tc\overline{\ts})$     &  0.283 & $\, \chi_{b0}(\tb\overline{\tb}) $ & 0.088   \\
\hline \hline
\end{tabular}
\end{center}
\end{table}
Our results reveal the following hierarchy among the 
$T_H$  temperatures,
\bea
\nn &&T_H^{\sigma} <T_H^{K_0^{*}}<T_H^{f_0}\,,\\
\nn &&T_H^{D_0^{*}} <T_H^{D_{s0}^{*}}\,,\\
\nn &&T_H^{B_{c0}} <T_H^{\chi_{c0}}<T_H^{\chi_{b0}}\,.
\eea
In the case of the  $B_0^{*}$ and  $B_{s0}$ the $T_H$ temperatures coincide.
Table \ref{table-mesones-desviaciones} shows the deviations of the screening masses at $T=500$ MeV from their free limits. In this S channel, the smallest deviation occurs for $\chi_{c0}(\tc\bar{\tc})$, with 9\%, while the largest deviation is observed for the meson $B_{0}^*(\tu\bar{\tb})$, which corresponds to a 57\%.
However, in the case of PS mesons, the screening masses exhibit a markedly different pattern, as will be discussed in the following section. At sufficiently high temperatures, it becomes possible to examine the connection between these states, offering a clear signature of chiral symmetry restoration. This feature shall be analyzed in the following sections.
\subsection{Pseudoscalar Mesons}
\begin{figure*}[ht]
\begin{tabular}{@{\extracolsep{-2.3 cm}}c}
 \renewcommand{\arraystretch}{-1.6} %
 \hspace{-1cm}
 \includegraphics[scale=0.6]{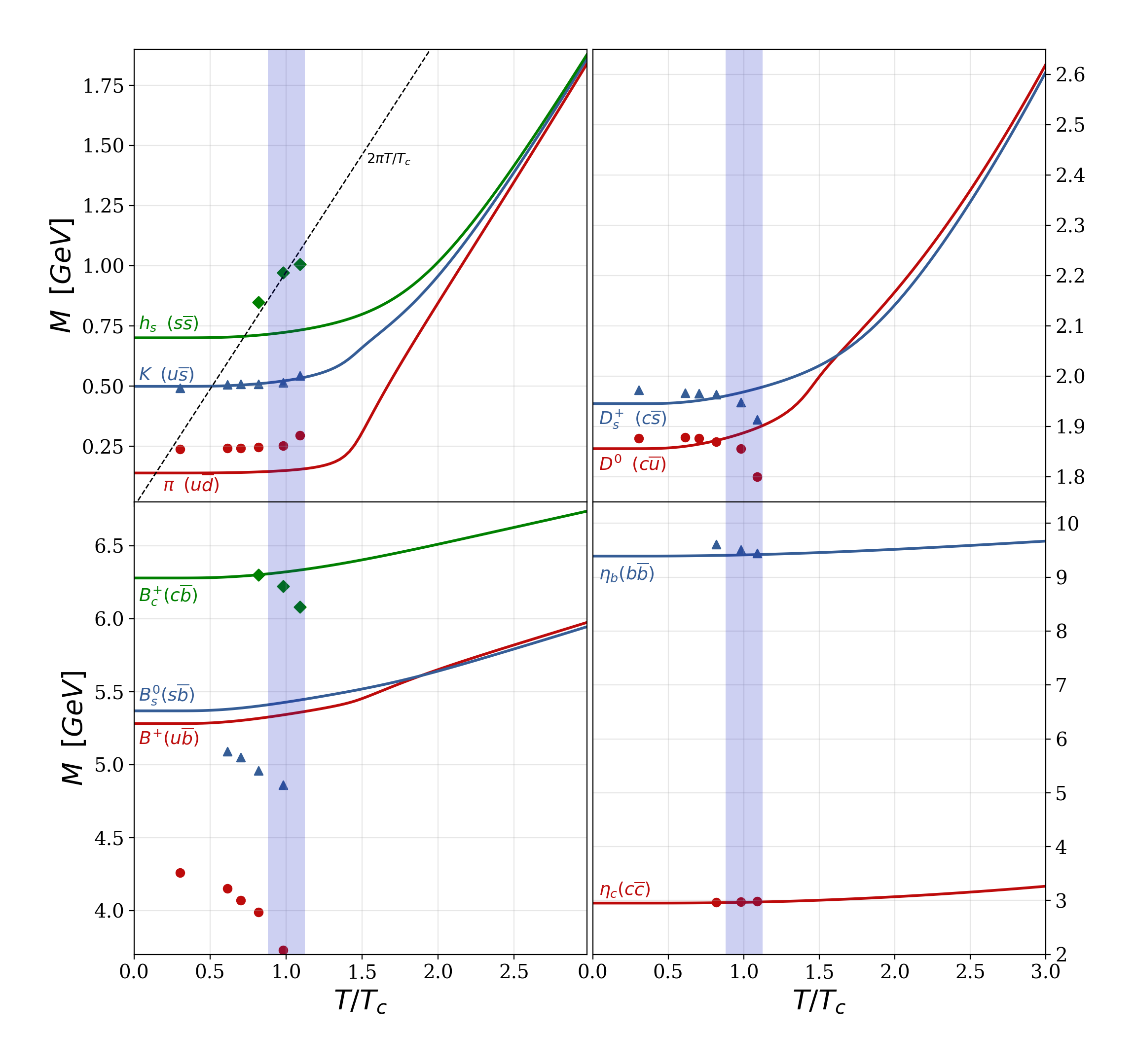}
\end{tabular}
\vspace{-1cm}
\caption{ \justifying \label{plotPS1}The screening mass for PS mesons composed of the five quark flavors ($\tu$, $\td$, $\ts$, $\tc$ and $\tb$). The dashed line correspond to the free theory limit of
$m=2\pi T$. The screening masses tend toward the mass of the meson at $T = 0$ MeV. As the temperature increases, the mass of the PS meson increases monotonically. This behavior aligns qualitatively with how the screening mass evolves with temperature in the DSE framework \cite{Blaschke:2000gd,Maris:2000ig,Wang:2013wk}. In the upper left  panel, the lightest mesons are shown; the upper right  panel displays mesons composed of one light quark and one charm quark. The lower  left  panel shows mesons containing different quarks, including a bottom quark, while the lower right panel presents mesons composed of two identical heavy quarks. Predictions obtained using other models are presented as markers preserving the color coding of our results in each panel for comparison; exact values are listed in Table~\ref{fboT}. }
\end{figure*}
We now focus on PS mesons, which are the lightest mesonic states and play a crucial role in the study of chiral dynamics. The BSA for a PS meson comprised of quarks with flavor $\fd$ and antiquarks with flavor $\fu$ at nonzero temperature is given by
\begin{equation}
\Gamma_{\Meps}(Q;T) = \gamma_5 \left[ i E_{\Meps}(T) + \frac{\gamma\cdot Q}{2M_R(T)}  F_{\Meps}(T) \right].
\end{equation}
and the corresponding kernel takes the form
\begin{align}
    \mathcal{K}_{\Meps}(Q;T)&=\Tr_{\rm D}\int\frac{d^4 q}{(2\pi)^4}
    \Gamma_{\Meps}(-Q;T)\frac{\partial}{\partial Q_\mu} S_{\fd}(q+Q;T)\nn \\ &\times \Gamma_{\Meps}(Q)S_{\fu}(q;T)
\end{align}
which explicitly takes the matrix form 
\begin{align}
\label{eigen}
    \left[ \begin{array}{c}
         E_{\Meps} (T) \\
          F_{\Meps} (T)
    \end{array}\right] = \frac{4\hat{\alpha}_{\rm IR}}{3\pi}
    \left[ \begin{array}{cc}
         \mathcal{K}_{\Meps}^{EE} (T) &\mathcal{K}_{\Meps}^{EF}(T) \\
          \mathcal{K}_{\Meps}^{FE} (T) & \mathcal{K}_{\Meps}^{FF}(T)
    \end{array}\right]
    \left[ \begin{array}{c}
         E_{\Meps}(T)  \\
          F_{\Meps}(T)
    \end{array}\right] \, 
\end{align}
where,
\begin{align*}
    \mathcal{K}_{\Meps}^{EE}(T) =&\int_0^1 d\alpha \,\bigg\{\mathcal{C}^\mathcal{T}\\
    +&\bigg[M_{\fu}(T)M_{\fd}(T)-\alpha(1-\alpha)Q^2-\omega^{(1)}(T)\bigg]\;\overline{\mathcal{C}}_1^\mathcal{T}\bigg\},\\
     \mathcal{K}_{\Meps}^{EF} (T)=&\frac{Q^2}{2M_R(T)}\int_0^1 d\alpha\bigg[(1-\alpha)M_{\fu}(T)+\alpha M_{\fd}(T)\bigg
]\overline{\mathcal{C}}_1^\mathcal{T},\\
\mathcal{K}_{\Meps}^{FE}(T) =&\frac{2M_R^2(T)}{Q^2}\mathcal{K}^{EF}_{\Meps}(T),\\
\mathcal{K}_{\Meps}^{FF}(T) =&-\frac{1}{2}\int_{0}^{1}\, d\alpha\bigg[M_{\fu}(T)M_{\fd}(T)\\
+&(1-\alpha)M_{\fu}^2(T)+\alpha M_{\fd}^2(T)\bigg]\overline{\mathcal{C}}_1^\mathcal{T} \, .
\end{align*}

\begin{figure}[h]
    \centering
    \includegraphics[width=1\linewidth]{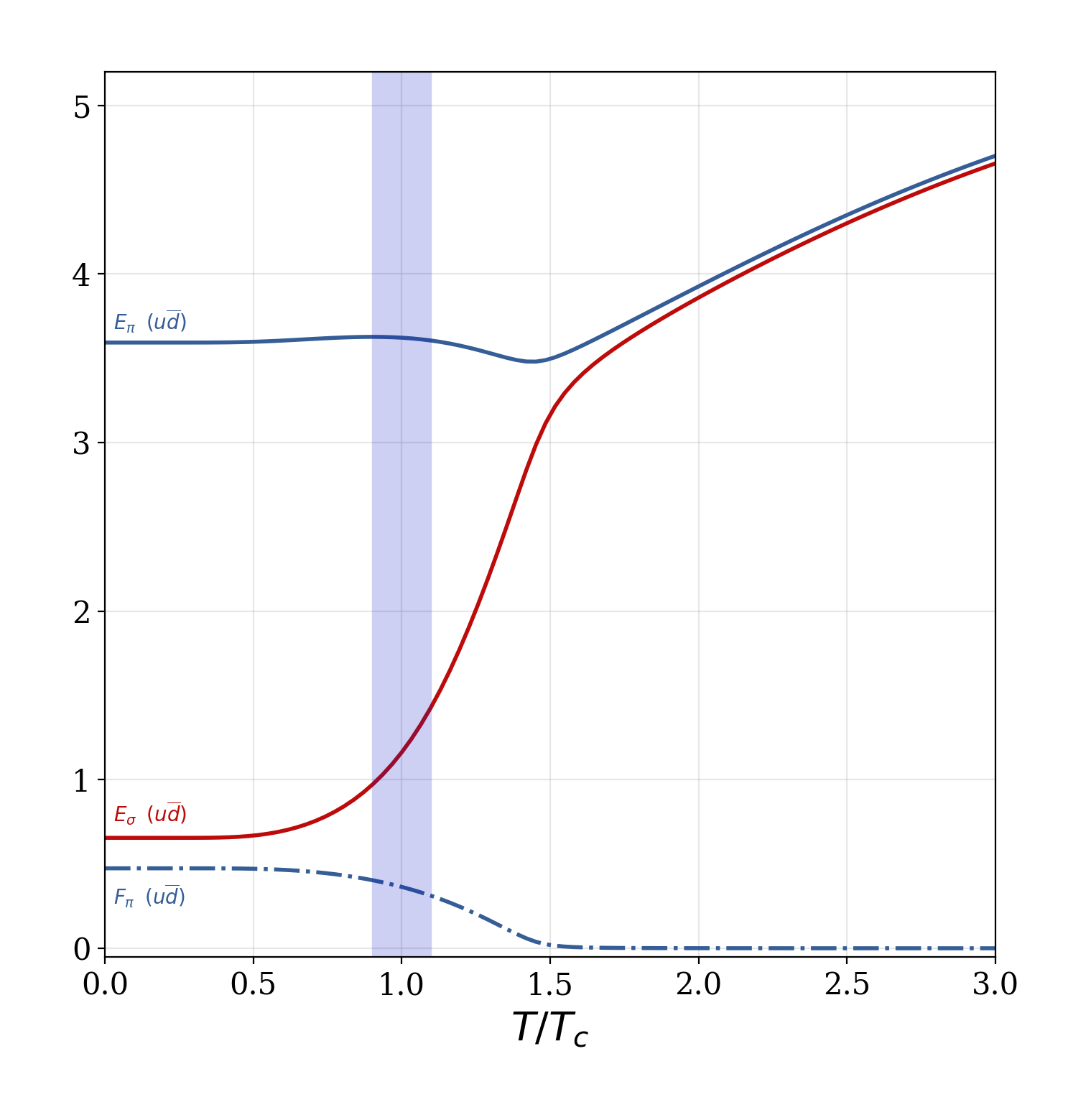}
    \vspace{-1cm}
    \caption{ \justifying BSA for the PS mesons  evaluated at  $Q=({\vec{Q},\omega_0})$ and plotted as function of $T$. The plot illustrates the convergence of $E_\pi$ and $E_\sigma$  to a common value at high temperatures. Moreover, due to the absence of an $F_\sigma$ component, the  $F_\pi$ component vanishes in the high-temperature regime.}
    \label{fig:ampPS}
\end{figure}

In the computation of observables, one must use the canonically-normalised BSA; i.e., $\Gamma_{\Meps}(Q;T)$ is rescaled so that
\begin{align}
\nn&\frac{Q_\mu}{N_c} =  \\
\nn&{\rm tr} \int\! \frac{d^4q}{(2\pi)^4}\Gamma_{\Meps}(-Q;T)
 \frac{\partial}{\partial Q_\mu} S(q+Q;T) \, \Gamma_{\Meps}(Q;T)\, S(q;T)\,. \label{Ndef}
\end{align}
In the chiral limit, this means
\begin{equation}
1 = \frac{N_c}{4\pi^2} \frac{1}{M^2(T)} \, {\cal C}^{\rm iu}_1(M^2(T);\tau_{\rm IR }^2,\tau_{\rm UV}^2)
E_{\Meps} [ E_{\Meps} - 2 F_{\Meps}].
\label{Norm0}
\end{equation}
With all these ingredients, the solution of \eqn{eigen}  provides the screening mass of the PS mesons  when $Q^2=-m_{\Meps}^2(T)$.
In Table \ref{table-mesones-pseudo}, we show our results at $T=0$ MeV for ten mesons and compare them with the experimental results, when possible. 
It is certain that at zero temperature we can recover the experimental results \cite{ParticleDataGroup:2024cfk}. 
Other studies using the CI model at $T=0$ \cite{Serna:2017nlr}, which only vary the parameter $\Lambda_{\rm UV}$
  for mesons containing the charm quark, yield similar mass values with percentage differences of less than 0.5\%.\\
\begin{table}[ht]
\caption{\justifying \label{table-mesones-pseudo}
Computed masses for PS mesons (GeV) and BSAs with the parameters of Tab.~\ref{parameters} and Tab.~\ref{table-M}. }
\begin{center}
\begin{tabular}{@{\extracolsep{0.2 cm}}cccccc}
\hline
\hline
Meson   & Exp. & CI & $E_{\Meps}$ & $F_{\Meps}$ & Charge \\ \hline
\rule{0ex}{2.5ex}
$\pi(\tu\bar{\td})$ &0.139&0.14&3.60&0.47&1\\
\rule{0ex}{2.5ex}
$K(\tu\bar{\ts})$ &0.493&0.49&3.81&0.59&1\\
\rule{0ex}{2.5ex}
$h_s(\ts\bar{\ts})$ & $\cdots$ & 0.69&4.04&0.74&0\\
\rule{0ex}{2.5ex}
$D^{0}(\tc\bar{\tu})$ &1.86& 1.87&3.03&0.37&0\\
\rule{0ex}{2.5ex}
$D^{+}_{\ts}(\tc\bar{\ts})$ &1.97 & 1.96& 3.24 & 0.51&1 \\
\rule{0ex}{2.5ex}
$B^{+} (\tu\bar{\tb})$ &5.28&5.28& 1.50&0.09&1 \\
\rule{0ex}{2.5ex}
$B_s^0(\ts\bar{\tb})$  &5.37& 5.37&1.59&0.13&0\\
\rule{0ex}{2.5ex}
$B_{\tc}^{+}(\tc\bar{\tb})$ &6.27&6.29&0.73& 0.11&1 \\
\rule{0ex}{2.5ex}
$\eta_c(\tc\bar{\tc})$ &2.98& 2.98&2.16& 0.41&0\\
\rule{0ex}{2.5ex}
$\eta_{\tb}(\tb\bar{\tb})$ &9.40& 9.40&0.48&0.10&0\\
\hline
\hline
\end{tabular}
\end{center}
\end{table}
Figure \ref{plotPS1} illustrates the temperature dependence of the screening masses over the range $T/T_c\in[0,300]$ MeV.
For improved clarity, we include a shaded purple band indicating the critical temperature range, $T_c\pm 0.1\,T_c$ GeV.
We compare our results with those obtained using lattice QCD calculations, Ref.\cite{Aarts:2022krz,Bazavov:2020teh,Mukherjee:2008tr}, within an efective theory approach \cite{Montana:2021vks}, in the basis of thermal QCD sum rules \cite{Veliev:2011zz} and in the framework of finite temperature QCD sum rules \cite{Veliev:2010vd}.
The results for the PS mesons are consistent with previous studies \cite{Chen:2024emt,Wang:2013wk}, which show that their screening masses increase with temperature. Below the critical temperature $T_c$, the increase is gradual, as can be observed at various temperatures in Tab.~\ref{fboT}. The temperatures below the critical point, namely $T=47,95,109,127$ and 
$152$ MeV, exhibit only modest growth, with percentage differences in the screening masses remaining below 12.65\%. However, once $T_c$ is surpassed around $T=169$ MeV, these differences increase up to 14.23\%, and at higher temperatures such as $T=500$ MeV, they reach approximately 92\%. The table also clearly shows that these changes are more pronounced for light mesons, while the increase in the screening masses of heavy mesons is less drastic. This behavior is observed not only for mesons composed of light quarks but also for those containing one or two heavy quarks. 
This highlights that dynamical chiral symmetry breaking and its restoration are less important effects in heavy-light mesons, and become almost negligible in pure bottomonium systems, where the quark masses primarily arise from the Higgs mechanism.
In most cases, our results align closely with those from other models. However, the limited availability of data for mesons with heavy quarks often leads various models using different approaches to reach differing conclusions.
Nevertheless, at very high temperatures ($T\to \infty$), these masses are expected to approach the value $2\sqrt{(\pi T)^2 +(2 M_R)^2 }$, independently of spin and flavor \cite{Detar:1987kae}. 
In Table \ref{table-mesones-desviaciones}, we show the deviation from this limit at $T=500$ MeV using our results.
\begin{table}[ht]
\caption{\justifying \label{table-mesones-desviaciones}
Screening masses in GeV at $T=500$ MeV, showing their deviation from the $2\sqrt{(\pi T)^2 +(2 M_R)^2 }$ limit at this temperature. The results marked with $\star$ correspond to the scenario where the quark masses are considered very small, that is, $M_R\to 0$, and the limit is $2\pi T$. }
\begin{center}
\begin{tabular}{lcc|lcc}
\toprule
\multicolumn{3}{c|}{Scalar} & \multicolumn{3}{c}{Pseudoscalar} \\
\hline
  & Mass  &\, Dev (\%) \, &  &  Mass  & \, Dev(\%) \, \\
  \hline
\rule{0ex}{2.5ex}
$\sigma(\tu\bar{\td})$ & 2.10 &42 & $\, \pi(\tu\overline{\td})$ & 2.09 & 42($33^\star$)\\
\rule{0ex}{2.5ex}
$K_0^*(\tu\bar{\ts})$ & 2.11 &42 &$\,  K(\tu\overline{\ts}) $ &2.11 &42($33^\star$)\\
\rule{0ex}{2.5ex}
$f_0(\ts\bar{\ts})$& 2.14 & 41&$\,  h_s(\ts\overline{\ts}) $ &2.12& 42($33^\star$)  \\
\rule{0ex}{2.5ex}
$D_0^*(\tc\bar{\tu})$& 2.68 &27 &$\,   D^{0}(\tc\overline{\tu})$ & 2.74 &25\\
\rule{0ex}{2.5ex}
$D_{s0}^*(\tc\bar{\ts})$ & 2.74 &27 &$\,  D_{\ts}^{+}(\tc\overline{\ts})$&2.73 &27  \\
\rule{0ex}{2.5ex}
$B_{0}^*(\tu\bar{\tb})$ & 5.84 & 57&$\, B^{+}(\tu\overline{\tb})$ &6.05 &63 \\
\rule{0ex}{2.5ex}
$B_{s0}(\ts\bar{\tb})$& 5.92 &54 &$\, B_{s}^{0}(\ts\overline{\tb})$ &6.02 &57 \\
\rule{0ex}{2.5ex}
$B_{c0}(\tc\bar{\tb})$ & 7.06 &47&$\, B_{\tc}^{+}(\tc\overline{\tb})$ & 6.79 &42 \\
\rule{0ex}{2.5ex}
$\chi_{c0}(\tc\bar{\tc})$ & 3.77 &9&$\, \eta_{c}(\tc\overline{\tc}) $ &3.32 &21 \\
\rule{0ex}{2.5ex}
$\chi_{b0}(\tb\bar{\tb})$ & 9.91 &41&$\, \eta_{\tb}(\tb\overline{\tb}) $ &9.70 & 39 \\
\hline
\hline
\end{tabular}
\end{center}
\end{table}
From the table, it is straightforward to see that the largest deviation occurs for the PS meson $\, B^{+}(\tu\overline{\tb})$,
while in this channel the smallest deviation is observed for $\, \eta_{c}(\tc\overline{\tc}) $
 with only a 21\% difference from the free limit. We also observe that, as $m_{\tu}\to 0$, the deviation decreases.
 The screening masses of the heavier PS mesons are significantly larger than the free theory value, which is attributed to the truncations present in our model. To approach this limit in the case of heavy mesons, temperatures higher than 500 MeV are required.
The results obtained at $T=500$ MeV allow us to observe the following mass splittings between mesons
\bea
\nn m_{\sigma}(\tu\bar{\td})-  m_{\pi}(\tu\overline{\td})&=&0.01   \, {\rm GeV},\\
\nn m_{K_0^*}(\tu\bar{\ts})-m_K(\tu\overline{\ts}) &=& 0 \,{\rm MeV},\\
\nn m_{f_0}(\ts\bar{\ts})-m_{h_s}(\ts\overline{\ts})&=&0.02 \,{\rm GeV},\\
\nn m_{D_0^*}(\tc\bar{\tu})- m_{D^{0}}(\tc\overline{\tu}) &=&-0.06 \,{\rm GeV},\\
\nn m_{D_{s0}^*}(\tc\bar{\ts}) -m_{D_{\ts}^{+}}(\tc\overline{\ts})&=& 0.01 \,{\rm GeV},\\
\nn m_{B_{0}^*}(\tu\bar{\tb})- m_{B^{+}}(\tu\overline{\tb}) &=&-0.21 \,{\rm GeV},\\
\nn m_{B_{s0}}(\ts\bar{\tb})-m_{B_{s}^{0}}(\ts\overline{\tb})&=& -0.1 \,{\rm GeV},\\
\nn m_{B_{c0}}(\tc\bar{\tb})-\, m_{B_{\tc}^{+}}(\tc\overline{\tb}) &=&0.27 \,{\rm GeV},\\
\nn m_{\chi_{c0}}(\tc\bar{\tc}) - m_{\eta_{c}}(\tc\overline{\tc}) &=&0.45 \,{\rm GeV},\\
m_{\chi_{b0}}(\tb\bar{\tb}) -m_{ \eta_{\tb}}(\tb\overline{\tb}) &=&0.21 \,{\rm GeV}.
\label{rest}\eea
From these equations, it is straightforward to see that the masses of the parity partners are nearly equal at this temperature. For example, in the light sector, the mass differences were approximately 750 MeV at zero temperature; however, at $T=500$ MeV the differences are reduced to less than 2\%.
Therefore, chiral restoration occurs at approximately 500 MeV, as evidenced by \eqn{rest}.\\
The screening mass associated with the pion is of particular interest, as this meson serves as the Goldstone boson of chiral symmetry. Our results indicate that the pion mass increases monotonically with temperature, in agreement with previous studies based on the Dyson–Schwinger and Bethe–Salpeter equations framework \cite{Gao:2020hwo}. 
Similarly as the S mesons, we can define a temperature ($T_H$) at which the screening mass suddenly changes its behavior. 
A characteristic temperature $T_H$ can be defined as the point at which the screening masses exhibit an inflection in their temperature dependence, marking the onset of accelerated growth. This is identified by the condition $M''(T_H)=0$, where the second derivative of the mass with respect to temperature vanishes.
For light mesons, this temperature corresponds to the critical temperature. We show this value for the different mesons in Tab. \ref{temp2}.
\begin{table}[t]
 \caption{ \justifying \label{temps pseudos} 
$T_H$ for the PS mesons. In this case $T_H$ marks the point where the initially monotonic increase in mass becomes significantly more rapid.} 
\begin{center}
\label{temp2}
\begin{tabular}{@{\extracolsep{0.0 cm}}  c | c | c | c }
\hline \hline
 \, Meson \, &\,  $T_{H}$ \, & \, Meson \, &\,  $T_{H}$ \, 
 \\
 \hline
 \rule{0ex}{2.5ex}
$\, \pi(\tu\overline{\td})$ & 0.182 & $\, B^{+}(\tu\overline{\tb}) $ &\, 0.061 \, \\ 
\rule{0ex}{2.5ex}
$\,  K(\tu\overline{\ts}) $ & \, 0.127 \, & $\, B_{s}^{0}(\ts\overline{\tb}) $ & 0.063 \\ 
\rule{0ex}{2.5ex}
$\,  h_s(\ts\overline{\ts}) $ & \, 0.118 \, & $\, B_{\tc}^{+}(\tc\overline{\tb}) $ &0.068 \\ 
\rule{0ex}{2.5ex}
$\,   D^{0}(\tc\overline{\tu})$ & \, 0.086 \, & $\, \eta_{c}(\tc\overline{\tc}) $ & 0.111 \\ 
\rule{0ex}{2.5ex}
$\,  D_{\ts}^{+}(\tc\overline{\ts})$     &  0.095 & $\, \eta_{\tb}(\tb\overline{\tb}) $ & 0.076 \\
\hline \hline
\end{tabular}
\end{center}
\end{table}
In addition, our results reveal the following hierarchy among the 
$T_H$  temperatures,
\bea
\nn &&T_H^{\pi} >T_H^{K}>T_H^{h_s}\,,\\
\nn &&T_H^{\eta_{\tc}} >T_H^{\eta_{\tb}}\,.
\eea
\begin{table*}[htp]
\caption{\justifying \label{fboT}The PS meson screening masses computed at $T=47,95,109,127,152,169$ MeV and $T=500$ MeV are compared with the lattice results presented in Ref.\cite{Aarts:2022krz,Bazavov:2020teh,Mukherjee:2008tr}, within an efective theory approach \cite{Montana:2021vks}, in the basis of thermal QCD sum rules \cite{Veliev:2011zz} and in the framework of finite temperature QCD sum rules \cite{Veliev:2010vd}. These temperatures span both the hadronic phase and the deconfined phase, lying below and above the critical temperature. The results indicate that the ground-state screening mass remains approximately constant at low temperatures, but exhibits a rapid increase once the critical temperature is exceeded. Columns four, six, eight, ten, twelve, and fourteen display the percentage difference between the screening masses in the adjacent columns. All quantities are in MeV. } 
\vspace{0 cm}
\begin{tabular}{@{\extracolsep{0.24cm}}c|cccccccccccccc}
\hline
\hline
 & $T$ (MeV) & 47 &  \% & 95 &   \% & 109 &   \%& 127 &  \%& 152 &  \% & 169& \%& 500\\
\hline
 \rule{0ex}{3.0ex}
 \multirow{2}{1.0cm}{$\pi$ } &CI & 139 &0.71 & 140 & 0.71& 141 & 2.08& 144 &3.35& 149 & 3.24& 154 & 92.64 & 2093
\\ \rule{0ex}{3.0ex}
& Ref. \cite{Aarts:2022krz} & 239 &1.24 &242&0.41 &243&1.22 & 246 & 2.76&253 &14.23& 295&$\cdots$ & $\cdots$ \\
\hline
 \rule{0ex}{3.0ex}
 \multirow{2}{1.0cm}{$K$ } &CI & 499 & 0.60& 502 &0.40 & 504&1.18 & 510 &2.11& 521 &2.25& 533 &74.74 & 2110
\\ \rule{0ex}{3.0ex}
& Ref. \cite{Aarts:2022krz} & 493 &2.76 & 507 & 0.39&509 &0.20& 508 &1.55&516&4.8 & 543 & $\cdots$ &$\cdots$  \\
\hline
 \rule{0ex}{3.0ex}
 \multirow{2}{1.0cm}{$h_{\ts}$ } &CI & 701 &0.28 & 703 &0.42 & 706 & 0.84& 712 &1.52& 723 & 1.22&732 & 65.54&2124
\\ \rule{0ex}{3.0ex}
& Ref. \cite{Bazavov:2020teh} & $\cdots$  &$\cdots$ & $\cdots$  &$\cdots$ & $\cdots$ & $\cdots$ & 849 & 12.65 & 972 & 3.37& 1006 & $\cdots$& $\cdots$ \\
\hline
 \rule{0ex}{3.0ex}
 \multirow{2}{1.0cm}{$D$ } &CI & 1855 &0.27 & 1860 &0.20 & 1864 &0.37&  1871 &0.74& 1885 &0.63& 1897 &30.9& 2746
\\ \rule{0ex}{3.0ex}
& Ref. \cite{Aarts:2022krz} & 1876 &0.11 & 1878 &0.11&1876 &0.37& 1869&0.70 & 1856 &3.11& 1800& $\cdots$ & $\cdots$\\
\hline
\rule{0ex}{3.0ex}
 \multirow{2}{1.0cm}{$D_{\ts}$ } & CI & 1945 & 0.15& 1948 &0.15& 1951 &0.26& 1956&0.56 & 1967&0.40 & 1975&27.76 & 2734
\\ \rule{0ex}{3.0ex}
& Ref. \cite{Aarts:2022krz} & 1972& 0.31& 1966 &0.05& 1965 &0.10& 1963&0.77 & 1948 &1.83& 1913 & $\cdots$ \\
\hline
 \rule{0ex}{3.0ex}
 \multirow{2}{1.0cm}{$B^+$ } &CI & 5281&0.23 & 5293 &0.17 & 5302&0.26 & 5316 &0.45& 5340 &0.34& 5358 &11.50& 6054
\\ \rule{0ex}{3.0ex}
& Ref. \cite{Montana:2021vks}& 4260 &2.65& 4150&1.97& 4070&2.01& 3990 &6.97& 3730 &$\cdots$& $\cdots$ & $\cdots$ &$\cdots$\\
\hline
 \rule{0ex}{3.0ex}
 \multirow{2}{1.0cm}{$B_{\ts}^0$ } &CI & 5368 & 0.19 & 5378&0.17 & 5387&0.26 & 5401&0.42 & 5424 &0.33& 5442&9.68 & 6025
\\ \rule{0ex}{3.0ex}
& Ref. \cite{Montana:2021vks} & 5160 &1.38& 5090&0.79& 5050&1.81& 4960 &2.06& 4860 & $\cdots$ & $\cdots$ & $\cdots$& $\cdots$ \\
\hline
 \rule{0ex}{3.0ex}
 \multirow{2}{1.0cm}{$B_{\tc}^+$ } &CI & 6277 &0.11& 6284 &0.10 & 6290 &0.16& 6300 &0.27& 6317 &0.24& 6332&6.77 & 6792
\\ \rule{0ex}{3.0ex}
& Ref. \cite{Veliev:2011zz} & $\cdots$&$\cdots$ &$\cdots$  &$\cdots$ &$\cdots$ & $\cdots$ & 6300 &1.29 & 6220&2.30 & 6080 &$\cdots$&$\cdots$\\
\hline
 \rule{0ex}{3.0ex}
 \multirow{2}{1.0cm}{$\eta_{\tc}$ } &CI & 2952 &0.03 & 2953&0.07 & 2955&0.10 & 2958 &0.24& 2965&0.20 & 2971&10.57 & 3322
\\ \rule{0ex}{3.0ex}
& Ref. \cite{Mukherjee:2008tr,Bazavov:2020teh} &$\cdots$ &$\cdots$  &$\cdots$ &$\cdots$ & $\cdots$ &$\cdots$ & 2970 &0.34 & 2980&0.33 & 2990 &21.93& 3830 \\
 \hline
 \rule{0ex}{3.0ex}
 \multirow{2}{1.0cm} {$\eta_{\tb}$ } &CI & 9392&0.04 & 9396&0.02 & 9398&0.05 & 9403&0.10 & 9412 &0.08& 9420&2.97 & 9708
\\ \rule{0ex}{3.0ex}
& Ref. \cite{Veliev:2010vd} &$\cdots$ &$\cdots$  &$\cdots$ &$\cdots$ & $\cdots$ & $\cdots$ & 9610 &1.05 & 9510&0.63 & 9450 & $\cdots$& $\cdots$\\
\hline
\hline
\end{tabular}
\end{table*}
On the other hand, we obtain the bound state amplitudes from the homogeneous BSE. Their evolution in $T$ is depicted in Figure \ref{fig:ampPS}.
 From this Fig. \ref{fig:ampPS} several key features can be identified. At $T=0$ MeV, the amplitudes correspond to mesons previously reported in the literature and are consistent with experimental values. As temperature increases, the behavior of the amplitudes closely mirrors that of the screening masses in both the PS and S channels. Above $1.8Tc$, the difference between the amplitudes $E_\pi$ and $E_\sigma$ 
  drops below 1.2\%, signaling that the lightest opposite-parity meson channels become nearly degenerate, which is a clear indication of the onset of symmetry restoration. Furthermore, we observe that for PS mesons the $F_\pi$ component vanishes at high temperatures, which is in perfect agreement with studies using the Schwinger–Dyson equations \cite{Maris:2000ig}.
%
\subsection{Diquarks}
\begin{figure*}[ht]
\begin{tabular}{@{\extracolsep{-2.3 cm}}c}
 \renewcommand{\arraystretch}{-1.6} %
 \hspace{-1cm}
 \includegraphics[scale=0.6]{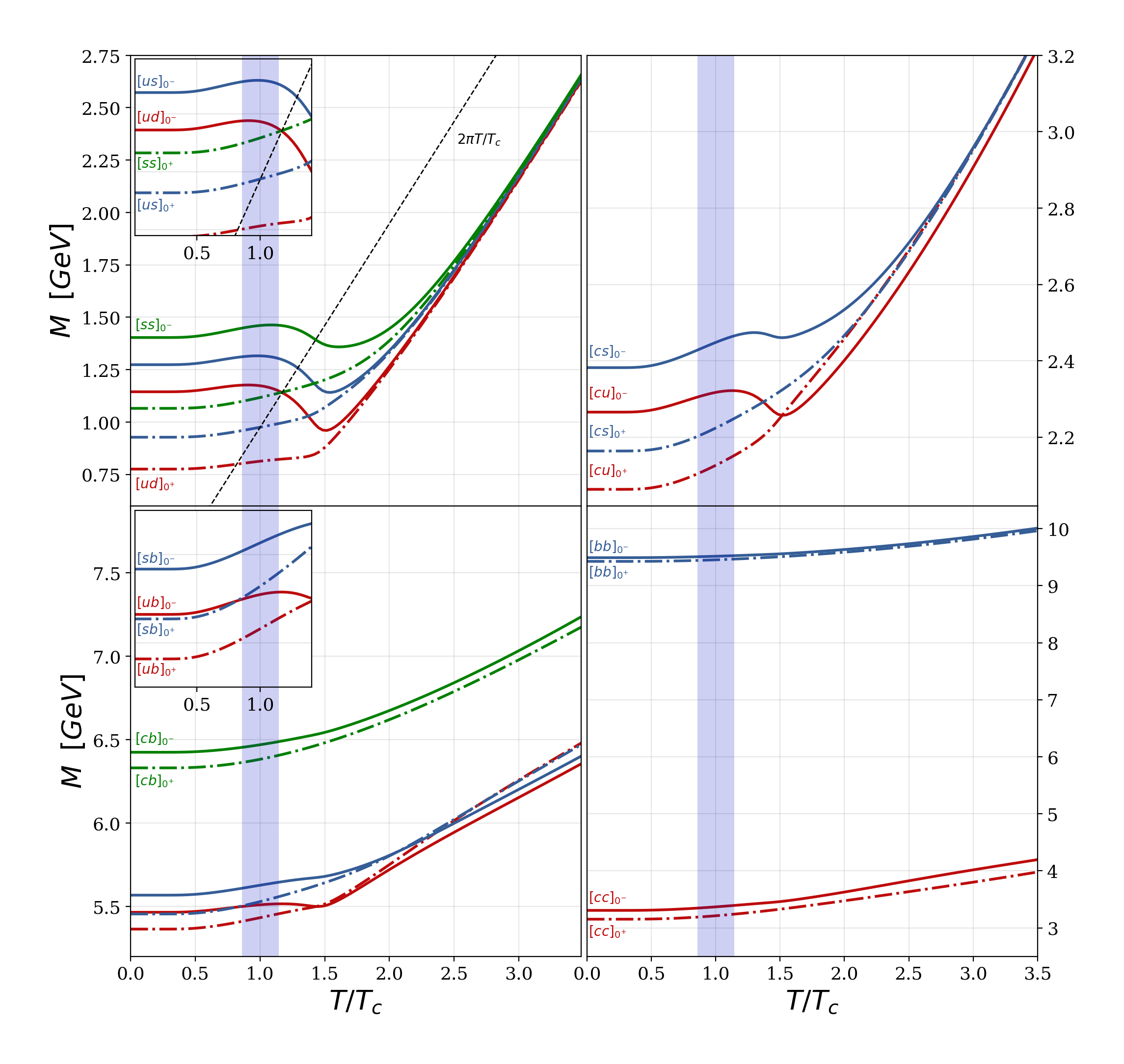}
\end{tabular}
\vspace{-1cm}
\caption{\justifying \label{plotdi} Comparison of the screening masses of PS and S diquarks.The shaded purple band indicates the critical temperature $T_c \pm 0.1\, T_c$. The solid lines correspond to  PS  diquarks, while the dashed-dotted lines represent  S diquarks.}
\end{figure*}
\begin{figure*}[ht]
\centering
\begin{subfigure}[b]{0.46\linewidth}
 \centering
\includegraphics[width=\linewidth]{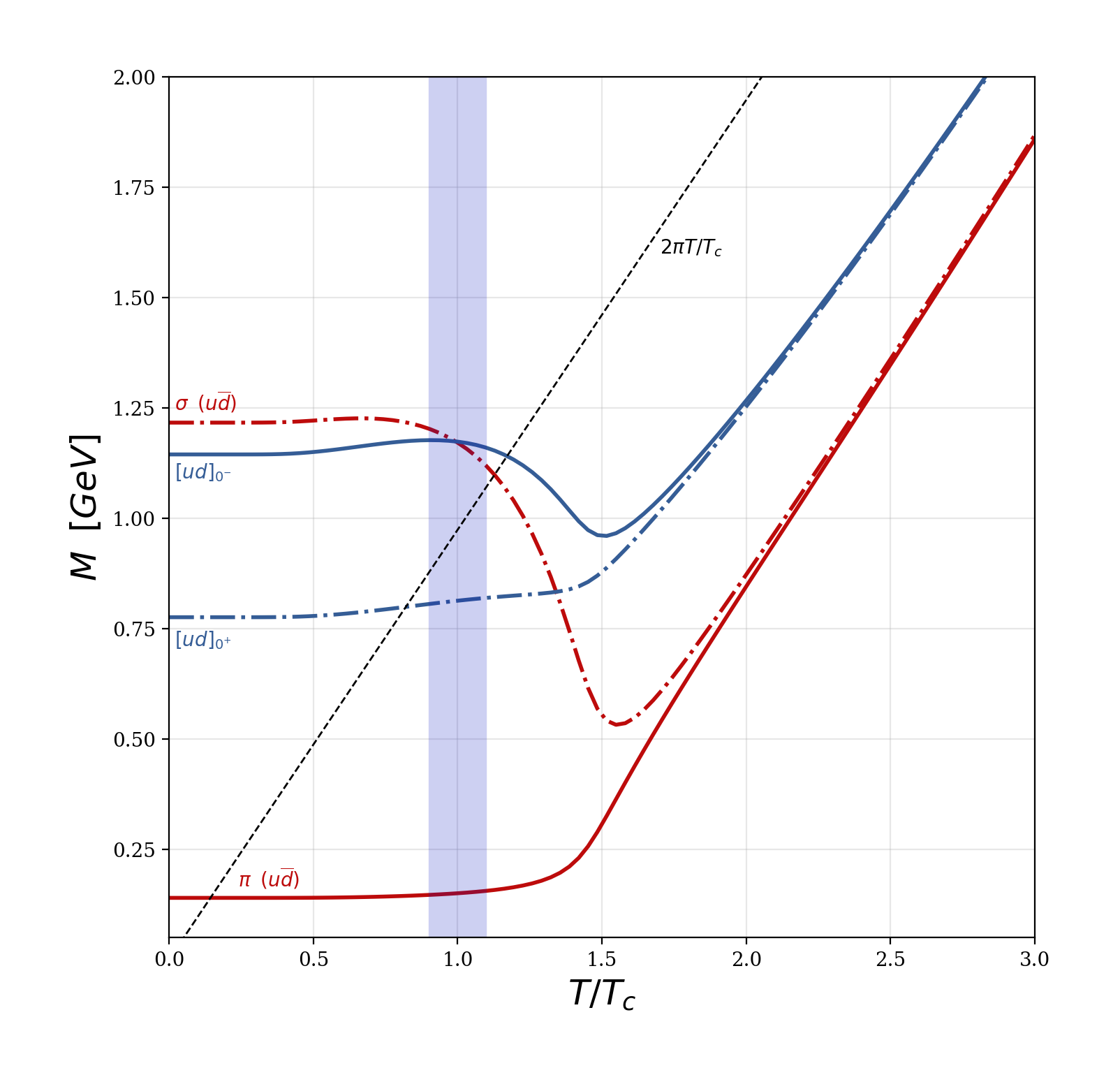}
\caption{Lightest mesons and diquarks}
\label{fig:ligeros}
\end{subfigure}
\begin{subfigure}[b]{0.46\linewidth}
 \centering
\includegraphics[width=\linewidth]{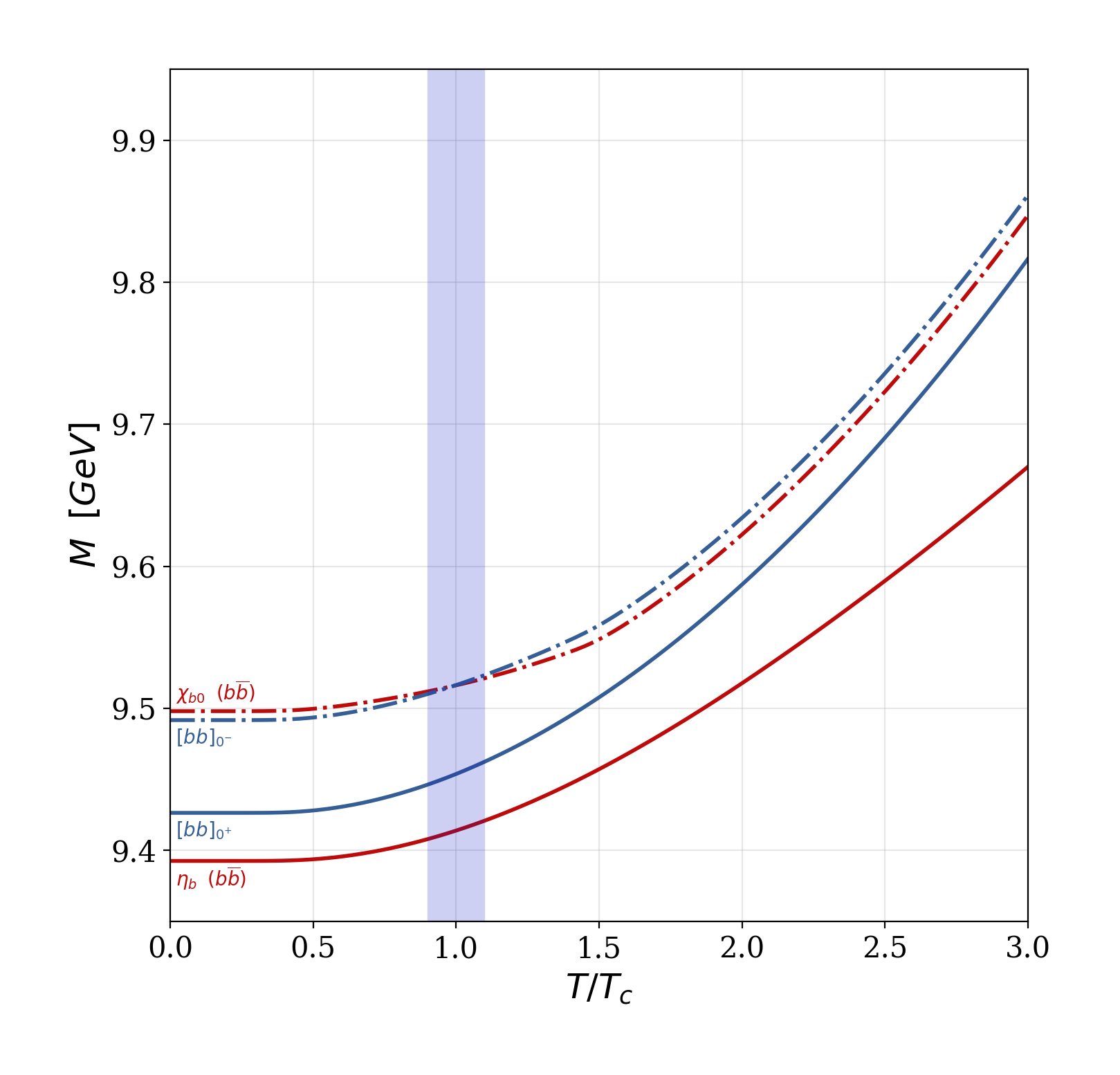}
\caption{Heaviest mesons and diquarks}
\label{fig:pesados}
\end{subfigure}
\caption{\justifying Comparison of the screening masses for the lightest and heaviest mesons and diquarks in the PS and S channels. Red lines denote mesonic states, while blue lines correspond to diquark states; dash-dotted curves represent positive-parity channels, and solid curves represent negative-parity channels.}
\label{fig:diyme}
\end{figure*}

An interaction that binds mesons also generates strong diquark correlations in the colour-$\bar 3$ channel.  It follows moreover that one may obtain the mass and Bethe-Salpeter amplitude for a diquark with spin-parity $J^P$ from the equation for a $J^{-P}$-meson in which the only change is a halving of the interaction strength.  The sign flip in parity arises from the fact that fermions and antifermions have opposite parity.
With this observation, we make the following changes to obtain the S and PS diquarks masses.
The eigenvalue equation in the case of scalar diquarks (DS) is given by
\begin{equation}
\label{bsefinalE}
\left[
\begin{array}{c}
E_{\Ds}(T)\\
\rule{0ex}{3.0ex} 
F_{\Ds}(T)
\end{array}
\right]
= \frac{4 \rmh}{6\pi}
\left[
\begin{array}{cc}
{\cal K}_{EE}^{\Meps}(T) & {\cal K}_{EF}^{\Meps} (T) \\
\rule{0ex}{3.0ex} 
{\cal K}_{FE}^{\Meps}(T)& {\cal K}_{FF}^{\Meps}(T)
\end{array}\right]
\left[\begin{array}{c}
E_{\Ds}(T)\\
\rule{0ex}{3.0ex} 
F_{\Ds}(T)
\end{array}
\right].
\end{equation}
Therefore, the equations that will provide us the masses of the PS diquarks (DPS)  is
\bea
\nn 0 & = & 1 + \frac{1}{2}{\cal K}_{\Ms}(-m_{\Deps}^2;T) \,. \\
\label{dq}\eea
It is worth emphasizing that a different value of $g_{\rm SO}^{qq}(T)$
  is employed for diquarks compared to mesons. This modification is essential to accurately reproduce the observed baryon masses, as detailed in Refs.\,\cite{Lu:2017cln,Yin:2021uom}. Specifically, the  used in the diquark sector is scaled by a factor of $1.8$ relative to that used for mesons. This enhancement effectively lowers the diquark masses, thereby allowing for a realistic description of the baryon spectrum. Accordingly, for the diquark sector we use the following  values,
\begin{eqnarray}
\label{gsodq}
g_{\rm SO}^{qq,0^-}(T) &=& 1-\frac{M_u(T)}{M_u(0)}(1-{0.58}^2)\,,
\end{eqnarray}
such as we recover at $T=0$ MeV, the spin-orbit enhacement predicted in previous analyses.

In Table~\ref{fbod}, we present the results for the diquarks obtained at $T=0$ MeV, which are in excellent agreement with those reported in Ref.~\cite{Gutierrez-Guerrero:2021rsx} and were used to calculate baryon masses. This fact implies that the computation of baryons at finite temperature within the quark-diquark framework will require the inputs calculated in this work.
\begin{table}[H] 
    \centering
    \caption{\justifying \label{fbod} Masses at $T=0$ MeV for scalar diquarks ($DS$) and pseudoscalar diquarks ($DPS$) obtained using the parameters described in Tab. \ref{parameters} and Tab. \ref{table-M}.}
    \begin{tabular}{@{\extracolsep{0.1 cm}} ccc|cc|ccc|cc}
         \hline
        \hline
        & &  & $DS$ & $DPS$ & &  & & $DS$ & $DPS$  \\  
         \hline
       \multirow{1}{*} & {$\tu\td$} & & $0.77$ & $1.15$ & & {$\tu\tb$} &  & $5.37$ & $5.47$  \\    
        \rule{0ex}{2.5ex}
         \multirow{1}{*} & {$\tu\ts$} &  & $0.92$ &  $1.27$ & & {$\ts\tb$} &  & $5.46$ &  $5.57$ \\
          \rule{0ex}{2.5ex}
         \multirow{1}{*}& {$\ts\ts$}  &  & $1.06$ &  $1.40$ & & {$\tc\tc$} & & $3.17$ &  $3.33$\\
          \rule{0ex}{2.5ex}
        \multirow{1}{*}& {$\tc\tu$} & & $2.08$ &  $2.28$  & & {$\tc\tb$} & & $6.35$ &  $6.44$  \\
         \rule{0ex}{2.5ex}
         \multirow{1}{*}& {$\tc\ts$} &  & $2.17$ &  $2.40$ & & {$\tb\tb$} &  & $9.43$ &  $9.50$\\
        \hline
        \hline
    \end{tabular}
\end{table}
In Figure \ref{plotdi}, we show the screening masses of the S and PS diquarks as a function of temperature. From this plot, it is immediately evident that S diquarks exhibit the same increasing behavior as PS mesons. Similarly, PS diquarks display a thermal behavior comparable to that observed in S mesons.
It is also evident that, as the temperature increases, the screening masses of diquark parity partners tend to converge.
To provide a clearer comparison of the behavior of mesons and diquarks, Fig.~\ref{fig:diyme} presents the screening masses of both the lightest and heaviest states in each channel.
This plot highlights key features of our results. For instance, the masses of the lightest parity partners converge to the same value, indicating
\begin{eqnarray}
\begin{array}{cc}
m_\pi\approx m_\sigma, & {\mbox{ when }\,\,\,\,} T>1.85 \,\, T_c\, ,  \\
m_{[ud]_{0-}}\approx m_{[ud]_{0+}} & {\mbox{ when }\,\,\,\,} T> 1.68 \,\, T_c \, .
\end{array}
\end{eqnarray}
In our study, for $T>1.85\,T_c$, the  mass difference between the pion and $\sigma$ mesons falls  below 3\%, indicating that the parity partners become effectively degenerate. This suggests that the dynamical chiral symmetry breaking effect weakens progressively as the temperature rises, and the chiral symmetry can be restored once the temperature reaches the critical value \cite{Mo:2010zza}.  Finally, in the heavy sector, the screening masses of mesons and diquarks exhibit a slower convergence toward their asymptotic values.
\section{ SUMMARY AND PERSPECTIVE}
\label{Summary}
In this work, we present the screening masses of spin-zero mesons and diquarks composed of light and heavy quarks—a total of forty states—using the Dyson-Schwinger equations (DSE) approach. We employ a symmetry-preserving treatment of a vector × vector contact interaction. Our results reveal several important features, which we summarize below:
\begin{itemize}
\item Our study of the $\tu,\td,\ts,\tc$ and $\tb$ quark masses reveals that, at high temperatures, their screening masses decrease, as shown in Figure~\ref{fig:dressed}. Our work also includes a comparison with other approaches, presented in Figure~\ref{fig:compa-njl}.
\item The S and PS meson masses at zero temperature are presented in Tables \ref{table-mesones-esc} and \ref{table-mesones-pseudo}, while the corresponding masses of their diquark partners at finite temperature are shown in Table \ref{fbod}. 
It is important to emphasize that our results for scalar mesons below 1$–$1.5 GeV correspond to hypothetical quark–antiquark bound states. These states are controversial due to the ongoing debate concerning their internal structure \cite{Santowsky:2020pwd,Santowsky:2021lyc,Santowsky:2021ugd}. In our analysis, they are included for completeness and in order to compute their corresponding diquark partners, which are essential for the calculation of baryon screening masses within the quark–diquark approximation \cite{Chen:2024emt}. In contrast, the heavier scalar mesons above this mass range, which have been experimentally measured, show good agreement with our predictions.
\item The behavior of the screening masses of mesons and diquarks are presented in Figures~\ref{fig:sscalar}, \ref{plotPS1} and \ref{fig:diyme}. In the calculations presented here, the screening masses of PS mesons exhibit a consistently increasing trend. In contrast, the S mesons show a markedly different behavior: their screening masses decrease up to a certain temperature $T_H$, and then increase again beyond that point. Our analysis indicates that the variation in behavior between pseudoscalar and scalar mesons is primarily due to the introduction of spin-orbit repulsion in the scalar channel, implemented through the phenomenological coupling  $g_{\rm SO}^{q\bar{q},0^+}(T)$.
This result is consistent with masses calculated using other models.
\item We also present in Table~\ref{table-mesones-desviaciones} the deviation of our screening masses for S and PS mesons from the free limit.
\item A clear correlation between the behavior of the screening masses and their corresponding Bethe–Salpeter amplitudes is evident, as illustrated in Figure \ref{fig:ampPS}.
    \item  In Figure~\ref{fig:diyme}, we show that  $m_\pi$ and $m_\sigma$ have approximately equal screening masses for $T>1.85\, T_c$,  providing clear evidence that chiral symmetry has been restored.
    \item The \eqn{rest} allows us to deduce that the masses of the meson parity partners become nearly degenerate at around 500 MeV; therefore, chiral restoration occurs at approximately this temperature for the mesons studied in this work.
    \item At sufficiently high temperatures, the screening masses of diquarks parity partners become  degenerate. Consequently, their contributions to the screening masses of baryons in the high-temperature regime become indistinguishable.
\end{itemize}
Our calculations are expected to provide valuable guidance for future experiments seeking to understand the fundamental aspects of the quark-gluon plasma in connection with heavy mesons.
An important contribution of our work is the study of heavy quarks as a function of temperature, as theoretical studies on this topic are scarce.
The investigations presented in this paper can be extended in several directions. A natural first step is to implement these calculations for vector and axial vector mesons. Given that the diquark results are already available, extending the framework to compute baryon properties at finite temperature, within the quark diquark picture, becomes a straightforward task. Moreover, the computation of meson elastic and transition form factors at nonzero temperature are both feasible and promising for future studies.

\begin{acknowledgements}
\vspace*{-2mm}
The authors would like to express their gratitude to the referee for their insightful comments.
L.~X.~Guti\'errez-Guerrero acknowledges the {\em Secretaría de Ciencia, Humanidades, Tecnología e Inovación} (SECIHTI) for the support provided to her through the {\em Investigadores e Investigadoras por México} SECIHTI program and Project CBF2023-2024-268, Hadronic Physics at JLab: Deciphering the Internal Structure of Mesons and Baryons, from the 2023-2024 frontier science call. The work of R.~J.~Hern\'andez-Pinto is partly supported by SECIHTI (Mexico) through {\em Sistema Nacional de Investigadores}. 
\end{acknowledgements}

\bibliography{ccc-a}

\end{document}